\documentclass[aps,prd,twocolumn,showpacs,showkeys,superscriptaddress,preprintnumbers,floatfix,reprint,nofootinbib]{revtex4}
\usepackage{graphicx}
\usepackage{dcolumn}
\usepackage{amsmath}
\usepackage{bm}
\usepackage{hyperref}
\usepackage{subfigure}
\usepackage{multirow}
\usepackage{subfigure}
\usepackage{rotating}
\usepackage{color}
\usepackage{multirow}
\usepackage{dcolumn}
\newcolumntype{C}[1]{>{\centering\arraybackslash}p{#1}}

\begin{document}

\preprint{
\vbox{
\hbox{ADP-16-29/T984}
}}

\def\sca{0.20}
\newcounter{rpl}
\renewcommand{\therpl}{\arabic{rpl}}
\newcommand{\rpl}{\refstepcounter{rpl}[(\therpl)]}

\title{Structure of the $\mathbf{\Lambda(1405)}$ from Hamiltonian effective field theory}

\affiliation{Centre for the Subatomic Structure of Matter (CSSM), Department of Physics, University of Adelaide, Adelaide SA 5005, Australia}
\affiliation{ARC Centre of Excellence for Particle Physics at the Terascale, Department of Physics, University of Adelaide, Adelaide SA 5005, Australia}

\author{Zhan-Wei Liu}
\affiliation{Centre for the Subatomic Structure of Matter (CSSM), Department of Physics, University of Adelaide, Adelaide SA 5005, Australia}

\author{Jonathan M. M. Hall}
\affiliation{Centre for the Subatomic Structure of Matter (CSSM), Department of Physics, University of Adelaide, Adelaide SA 5005, Australia}

\author{Derek B. Leinweber}
\affiliation{Centre for the Subatomic Structure of Matter (CSSM), Department of Physics, University of Adelaide, Adelaide SA 5005, Australia}

\author{Anthony W. Thomas}
\affiliation{Centre for the Subatomic Structure of Matter (CSSM), Department of Physics, University of Adelaide, Adelaide SA 5005, Australia}\affiliation{ARC Centre of Excellence for Particle Physics at the Terascale, Department of Physics, University of Adelaide, Adelaide SA 5005, Australia}

\author{Jia-Jun Wu}
\affiliation{Centre for the Subatomic Structure of Matter (CSSM), Department of Physics, University of Adelaide, Adelaide SA 5005, Australia}

\begin{abstract}
The pole structure of the $\Lambda(1405)$ is examined by fitting the couplings of an underlying
Hamiltonian effective field theory to cross sections of $K^- p$ scattering in the infinite-volume
limit. Finite-volume spectra are then obtained from the theory, and compared to lattice QCD results
for the mass of the $\Lambda(1405)$.  
Momentum-dependent, non-separable potentials motivated by the well-known Weinberg-Tomozawa terms
are used, with SU(3) flavour symmetry broken in the couplings and masses.
In addition, we examine the effect on the behaviour of the spectra from the inclusion of a bare
triquark-like isospin-zero basis state. It is found that the cross sections are consistent with the
experimental data with two complex poles for the $\Lambda(1405)$, regardless of whether a bare
baryon basis state is introduced or not.  However, it is apparent that the bare baryon is important
for describing the results of lattice QCD at high pion masses.
\end{abstract}

\maketitle


\section{Introduction}
Strange quark phenomenology has always been of great interest to both theoretical and experimental
physicists. It exhibits some properties of both light and heavy quarks. However, unlike either the
light quark limit or the heavy quark limit, it is difficult to explain the phenomena of strange
quark physics comprehensively simply by applying chiral symmetry for zero-mass quarks, and
heavy-quark symmetry for infinite-mass quarks.  Corrections to these symmetries and the treatment
of symmetry-breaking effects provide important insight into the internal structure of hadrons.

\subsection{The $\mathbf{\Lambda(1405)}$}
The $\Lambda(1405)$ is a resonant state with strangeness number $S=-1$, and
$I(J^P)=0(\frac12^-)$. It also has been shown to interact strongly with nearby meson-baryon states,
the details of which are intimately dependent on its internal structure. The $\Lambda(1405)$ is
close to the threshold of $\bar K N$ and mainly decays to the $\pi \Sigma$ state. With the
interactions of just these two channels in a Hamiltonian model, the data of $\bar K N$ scattering
at low energy can be fit well, and the $\Lambda(1405)$ resonance can be generated. Corrections from
the interaction with $\eta \Lambda$ and $K \Xi$ are also usually considered in studying this
problem.

A two-pole structure of the $\Lambda(1405)$ has been proposed in many works
\cite{Kaiser1997,Oset1998,Oller2001,Hyodo2004}. Many groups claim these two poles lie on one
Riemann sheet. Besides the traditional pole around $1420-25 i$ MeV, there is another pole with the
position varying widely in different models or in different fits
\cite{Ikeda2011,Guo2013,Mai2013}. The second pole is not mentioned in every work
\cite{Akaishi2008}.

\subsection{Three-quark Core Contributions}

The basis of the quark model is that there is a three-quark 
state which provides the dominant contribution to the properties 
of all baryons.
For example, the nucleon is considered to be dominated by a bare nucleon state dressed
by smaller contributions from $\pi N$, $\pi \Delta$, 
etc.~\cite{Thomas:1981vc,Thomas:1982kv}.
However, the discussion regarding the
$\Lambda(1405)$ remains ongoing, and deserves careful analysis.

Lattice QCD calculations are able to excite the $\Lambda(1405)$ with local three-quark operators
\cite{Menadue2012,Engel2013a,Hall2015} suggesting a nontrivial role for a three-quark component.
However, lattice QCD calculations of the strange magnetic form factor of the $\Lambda(1405)$ have
revealed the $\Lambda(1405)$ to be dominated by a molecular $\bar K N$ bound state at light quark
masses \cite{Hall2015,Hall2014}.  Attraction between the $\bar K$ and $N$ provides clustering which
avoids the common volume suppression of weak-scattering states.  A small bare-state component also
provides a mechanism for the excitation of this state with local three-quark operators.

As one varies the light quark mass in lattice QCD, one expects that the ``bare'', tri-quark state
will tend to be more important as chiral loops are suppressed in that
region~\cite{Young:2002cj,Hall2015,Hall2014}.  
For example, one can fit the experimental data very well in the absence of a bare state
contribution with the aforementioned two-particle channels~\cite{Molina2015}. Such an approach also
gives good predictions for the lattice results at small pion masses where the unitary chiral
extrapolation works.

However, in the high-pion-mass region the energies of the lattice QCD eigenstates in the finite
volume spectrum are much smaller than the thresholds of the reaction channels.
It is not possible to form a bound state with a binding energy of more than 100 MeV without a large
increase in the coupling parameters as the pion mass increases.
While this cannot be ruled out by our analysis, we consider the interpretation in terms of a bare
(three-quark) state to be more natural.

Identification of the $\Lambda(1405)$ in the finite volume of the lattice was performed
\cite{Hall2015,Hall2014} using a simple Hamiltonian effective field theory model focusing on the
flavour singlet couplings of the $\pi \Sigma$, $\bar K N$, $\eta \Lambda$ and $K \Xi$ channels to
the bare basis state required by the admission of a three-quark configuration carrying the quantum
numbers of the $\Lambda(1405)$.  Having established that the $\Lambda(1405)$ is a molecular $\bar K N$
bound state \cite{Hall2015,Hall2014} it is important to examine the finite volume spectrum in a
calculation that does not distinguish the flavour symmetry of the isospin-zero state.  One should not
only examine the $\Lambda(1405)$, but also examine the low-lying excitations observed in
Ref.~\cite{Menadue2012}, associated with the octet-flavour interpolating fields used in the
lattice correlation matrix.

\subsection{Models of the $\mathbf{\Lambda(1405)}$}

Many models have been applied to the $\Lambda(1405)$
\cite{Rajasekaran1972,Leinweber:1989hh,Kaiser1997,Oset1998,Oller2001,Hyodo2004,Akaishi2008,%
Hall2015,Hall2014,%
Hyodo2008,Ikeda2011,Ikeda2012,Guo2013,Mai2013,Mai2015,Molina2015,Cieply2016,%
Fernandez-Ramirez2016,Kamiya2016,Ohnishi2016,Jido:2003cb,Veit:1984an,Veit:1984jr,Cieply:2011nq}. 
Beginning with the study based on SU(3) chiral symmetry within the cloudy bag
model~\cite{Veit:1984an,Veit:1984jr}, Weinberg-Tomozawa terms have been successful in describing
the most prominent interactions. Dimensional regularisation was used in solving the Bethe-Salpeter
equation with the full Weinberg-Tomozawa potential in Refs.~\cite{Mai2013,Mai2015}.  It is also
common to use a $K$-matrix approach in which the potentials are used with an on-shell approximation
to obtain the scattering amplitude~\cite{Oset1998,Ikeda2011,Ikeda2012}. In that case, the potential
is often taken to be momentum independent. Regularisation with a cutoff was taken in
Refs.~\cite{Ohnishi2016,Akaishi2008}. Separable potentials are favoured since they are easy to
solve \cite{Akaishi2008}.

Rather than effective Weinberg-Tomozawa potential, hadron-exchange potentials are used to study $\Lambda(1405)$ \cite{MuellerGroeling:1990cw,Haidenbauer:2010ch}. This dynamical coupled-channel approach is also discretized to study the spectrum on the lattice \cite{Doring:2011ip}. 

In this work, we use Hamiltonian effective field theory to analyze both the available experimental data
in infinite volume and the results from lattice QCD at finite volume.
Hamiltonian effective field theory is a powerful tool for analyzing the lattice results and
examining the structure of the states on the lattice
\cite{Hall2013,Hall2014,Hall2015,Liu2016a,Wu2014,Leinweber2015}. Moreover, it can be applied to
calculate the scattering processes in infinite volume.
The Weinberg-Tomozawa potentials are included in a Hamiltonian model of the $\Lambda(1405)$, which
matches finite-volume effective field theory. The on-shell approximation is not used, and thus the
potentials are momentum dependent and non-separable. The effect of the bare baryon is carefully
examined by comparing the results of two scenarios, with and without a bare baryon basis state in
the formulation of the Hamiltonian matrix.  The two-pole structure of the $\Lambda(1405)$ is also
examined.

\subsection{Outline}

The formalism for our Hamiltonian effective field theory is presented in Sec.~\ref{sec:Frm}.  
In constructing the Hamiltonian, we consider two scenarios: one in which the $\Lambda(1405)$ is
dynamically generated purely from the $\pi \Sigma$, $\bar K N$, $\eta \Lambda$ and $K \Xi$
interactions, and one also including a bare-baryon basis state to accommodate a three-quark
configuration carrying the quantum numbers of the $\Lambda(1405)$.

The forms of the interactions are described in Sec.~\ref{subsec:Hml}.
We then proceed to solve the Bethe-Salpeter equation and obtain the cross sections
and pole positions at infinite volume via the $T$-matrix in Sec.~\ref{subsec:TMT}.
To compare with lattice QCD results, the Hamiltonian is discretised in Sec.~\ref{subsec:FVM} and
solved to obtain results at finite volume.  

In Sec.~\ref{sec:NRD}, the numerical results are presented.  The experimental data for $\bar K N$
scattering is fit in Sec.~\ref{subsec:ExpCSP} and the calculation is extended to varying quark
masses in Sec.~\ref{subsec:LattFV}.  Then the finite-volume spectrum of the Hamiltonian model in our two
scenarios is compared to lattice QCD results.  Sec.~\ref{sec:convAnalysis} presents results in
the absence of a bare-baryon basis state and Sec.~\ref{sec:BBanalysis} illustrates how the
inclusion of a bare-baryon basis state resolves discrepancies and provides an explanation of which
states are seen in contemporary lattice QCD calculations.
A brief summary concludes in Sec.~\ref{sec:sum}.

\section{Framework} \label{sec:Frm}

\subsection{Hamiltonian} \label{subsec:Hml}

To study the data relevant to the $\Lambda(1405)$, we consider the interactions among
$|\pi\Sigma\rangle$, $|\bar K N\rangle$, $|\eta\Lambda\rangle$, $|K\Xi \rangle$, and the isospin-1
channel $|\pi \Lambda\rangle$.  We use the following Hamiltonian to describe the interactions
\begin{eqnarray}
H^I = H^I_0 + H^I_{\rm int},
\label{eq:h}
\end{eqnarray}
where superscript $I$ is the isospin.

In the centre-of-mass frame, the kinetic-energy Hamiltonian $H^I_0$ is written as
\begin{eqnarray}
H^I_0 &=&\sum_{B_0} |B_0\rangle \, m^{0}_{B}  \, \langle B_0|+ \sum_{\alpha}\int d^3\vec{k}\nonumber\\
&&
|\alpha(\vec{k})\rangle\, \left[\, \omega_{\alpha_M}(k)+\omega_{\alpha_B}(k)\, \right]
\,\langle\alpha(\vec{k})| \, ,
\label{eq:h0}
\end{eqnarray}
where $|\alpha\rangle$=$|\pi\Sigma\rangle$, $|\bar K N\rangle ,\ \ldots$ 
and 
\begin{equation}
\omega_X(k)=\sqrt{m_X^2+k^2}
\end{equation}
is the non-interacting energy of particle $X$. The subscripts $\alpha_M$ and $\alpha_B$ represent
the meson and baryon separately in channel $\alpha$.

The interaction Hamiltonian of this system includes two parts
\begin{eqnarray}
H^I_{\rm int} = g^I + v^I.\label{eq:hi}
\end{eqnarray}
$g^I$ describes the vertex interaction between the bare baryon and two-particle channels $\alpha$
\begin{eqnarray}
g^I &=& \sum_{\alpha, B_0}\int d^3\vec{k} \, \left\{\,  |\alpha(\vec{k})\rangle \,
G^{I\dagger}_{\alpha, B_0}(k) \, \langle B_0| \right . \nonumber\\
&&\quad
+ \left . |B_0\rangle \, G^I_{\alpha, B_0}(k) \, \langle \alpha(\vec{k})|\, \right\},
\label{eq:int-g}
\end{eqnarray}
where we use the form of the ordinary S-wave coupling for $G^I$
\begin{equation}
G^I_{\alpha, B_0}(k)=\frac{\sqrt3\, g^I_{\alpha, B_0}}{2\pi f} \, \sqrt{\omega_\pi(k)} \,\, u(k).
\end{equation}
A dipole form factor, $u(k) = ( 1 + {k^2}/{\Lambda^2} )^{-2}$, with regulator parameter $\Lambda =
1$ GeV is used to regulate the calculation.
In the scenario without a bare baryon, we set the couplings $g^I_{\alpha, B_0}=0$ to turn off the
effect of $|B_0\rangle$.

The use of a dipole regulator has received a great deal of attention in the literature. 
It has been clearly established that this approach, known as finite-range regularization (FRR), is
equivalent to dimensionally regulated chiral perturbation theory ($\chi$PT) in the power counting
regime, \cite{Young:2002ib, Hall:2010ai} roughly below a 300 MeV pion mass, corresponding to the few
lowest lattice data points.
At higher pion masses the formal $\chi$PT expansion fails to converge. 
FRR provides a model for the behaviour of the chiral loops at larger meson mass which has proven
successful over a very wide range of pion masses for many observables.
By fitting the theory to both the experimental data and the lattice data, the other parameters in
the model acquire an implicit dependence on the regulator parameter, which removes the formal
dependence on that mass parameter.
In our work, all parameters and the bare state mass are appropriate to the regulator mass used, namely 1 GeV.

We define the direct two-to-two particle interaction $v^I$ by
\begin{eqnarray}
v^I = \sum_{\alpha,\beta} \int d^3\vec{k} \, d^3\vec{k}'\, |\alpha(\vec{k})\rangle \,
V^{I}_{\alpha,\beta}(k,k') \, \langle \beta(\vec{k}')|\, ,
\label{eq:int-v}
\end{eqnarray}
where we use the potential derived from the Weinberg-Tomozawa term \cite{Veit1985}
\begin{equation}
V^I_{\alpha,\beta}(k,k')=g_{\alpha,\beta}^I\frac{ \left [\, \omega_{\alpha_M}(k)+\omega_{\beta_M}(k')\,\right] \, u(k) \,u(k')}{8\pi^2 f^2 \,\sqrt{2\omega_{\alpha_M}(k)} \, \sqrt{2\omega_{\beta_M}(k')}}. \label{eq:vsp}
\end{equation}
We do not use the so-called on-shell approximation in this work. We keep the form
$\omega_{\alpha_M}(k)+\omega_{\beta_M}(k')$ rather than replacing it with
$2E-m_{\alpha_B}+m_{\beta_B}$ in Eq. (\ref{eq:vsp}). Our potentials are momentum dependent and not
separable.

\subsection{$\mathbf{T}$-Matrix}\label{subsec:TMT}

We can evaluate the $T$-matrices for two particle scattering 
by solving a three-dimensional reduction of the coupled-channel Bethe-Salpeter equation
in each partial wave
\begin{eqnarray}
&&T^I_{\alpha, \beta}(k,k';E)=\tilde V^I_{\alpha, \beta}(k,k';E)+\sum_\gamma \int q^2 \, dq\nonumber\\
&&\quad \tilde V^I_{\alpha, \gamma}(k,q;E) \, \frac{1}{E-\omega_\gamma(q)+i \epsilon} \,  T^I_{\gamma, \beta}(q,k';E),
\label{eq:BS}
\end{eqnarray}
where $\omega_\alpha(k)$ is the total kinetic energy of channel $\alpha$,
\begin{equation}
\omega_{\alpha}(k)=\sqrt{m_{\alpha_1}^2+k^2}+\sqrt{m_{\alpha_2}^2+k^2},
\end{equation}
and the coupled-channel potential can be obtained from the interaction Hamiltonian
\begin{eqnarray}
\tilde V^I_{\alpha, \beta}(k,k';E) &=& \sum_{B_0} G^{I\dag}_{\alpha, B_0}(k) \, \frac{1}{E-m_B^0} \, G^I_{\beta, B_0}(k') 
\nonumber\\&&\qquad
+V^I_{\alpha,\beta}(k,k').
\label{eq:lseq-2}
\end{eqnarray}
The cross section $\sigma_{\bar \alpha, \bar \beta}$ for the process $\bar \beta \to \bar \alpha$ is 
\begin{equation}
\sigma_{\bar \alpha, \bar \beta}=\frac{4\pi^3 \, k_{\rm cm}^\alpha \, \omega^{\rm cm}_{\alpha_M} \, \omega^{\rm cm}_{\alpha_B} \, \omega^{\rm cm}_{\beta_M} \, \omega^{\rm cm}_{\beta_M} 
}{ E_{\rm cm}^2 \,  k_{\rm cm}^\beta}  \, |T_{\bar \alpha, \bar \beta}(k_{\rm cm}^\alpha, k_{\rm cm}^\beta;E_{\rm cm})|^2,  \label{eq:sigma}
\end{equation}
where $T_{\bar \alpha, \bar \beta}$ is the linear combination of $T^0_{\alpha, \beta}$ and $T^1_{\alpha, \beta}$ multiplied by the corresponding Clebsch-Gordan coefficients, e.g. $T_{\bar K^0 n, K^- p}=-1/2 \, T^0_{\bar K N, \bar K N}+1/2 \, T^1_{\bar K N, \bar K N}$. The superscript and subscript ``cm'' refer to the center-of-mass frame.

To find the poles of $T^0_{\alpha, \beta}(k,k';E_{\rm pole})$, we replace the integration variable $q$ with $q\times \exp(-i\theta)$, for $\gamma=\pi\Sigma$ in Eq. (\ref{eq:BS}), and maintain $0\ll \theta<\pi/2$. That is, we search for poles of the $\Lambda(1405)$ on the second Riemann sheet, which is adjacent to the physical sheet separated by the cut between the $\pi\Sigma$ and $\bar K N$ thresholds.

\subsection{Finite-Volume Matrix Hamiltonian Model}\label{subsec:FVM}

We can discretise the Hamiltonian in a box with length $L$ for $I=0$. A particle can only carry
momenta $k_n=\sqrt{n} \, 2\pi / L$ in the box, where $n=0,1, ...$. The non-interacting isospin-zero
Hamiltonian can be written as
\begin{equation}
\mathcal H^0_0={\rm diag}\{m_B^0, \omega_{\pi\Sigma}(k_0),\omega_{\bar K
  N}(k_0),...,\omega_{\pi\Sigma}(k_1),...\}\, ,
\end{equation}
and the interacting Hamiltonian is 
\begin{widetext}
\begin{equation}
\mathcal H^0_{\rm int}=\left( \begin{array}{cccccc}
                         0 & \mathcal  G^0_{\pi\Sigma, B_0}(k_0) &                  \mathcal  G^0_{\bar K N, B_0}(k_0) & \ldots&                  \mathcal  G^0_{\pi\Sigma, B_0}(k_1) &  \ldots \\
  \mathcal  G^0_{\pi\Sigma, B_0}(k_0) &\mathcal  V^0_{\pi\Sigma,\pi\Sigma}(k_0,k_0)&\mathcal  V^0_{\pi\Sigma,\bar K N}(k_0,k_0) & \ldots&   \mathcal  V^0_{\pi\Sigma,\pi\Sigma}(k_0,k_1) &                                        \ldots \\
     \mathcal  G^0_{\bar K N, B_0}(k_0) & \mathcal  V^0_{\bar K N,\pi\Sigma}(k_0,k_0)&\mathcal  V^0_{\bar K N,\bar K N}(k_0,k_0) & \ldots&   \mathcal  V^0_{\bar K N,\pi\Sigma}(k_0,k_1)&                                        \ldots \\
        \vdots &  \vdots & \vdots &    \ddots &   \vdots &                                       \ddots \\
\mathcal  G^0_{\pi\Sigma, B_0}(k_1) &\mathcal  V^0_{\pi\Sigma,\pi\Sigma}(k_1,k_0)&\mathcal  V^0_{\pi\Sigma,\bar K N}(k_1,k_0) & \ldots&   \mathcal  V^0_{\pi\Sigma,\pi\Sigma}(k_1,k_1) &                                        \ldots \\
                    \vdots &                   \vdots &                                 \vdots &                                      \ddots&                         \vdots &                                   \ddots \\
\end{array}
\right) \, ,
\end{equation}
\end{widetext}
where
\begin{eqnarray}
&&\hspace{-3em}
\mathcal G^0_{\alpha,B_0}(k_n)=\sqrt{\frac{C_3(n)}{4\pi}}  \left(\frac{2\pi}{L}\right)^{3/2} \, G^0_{\alpha,B_0}(k_n),\\
&&\hspace{-3em}
\mathcal V^0_{\alpha,\beta}(k_n, k_m)=\frac{\sqrt{C_3(n)\,C_3(m)}}{4\pi} \left(\frac{2\pi}{L}\right)^{3} \,
  V^0_{\alpha,\beta}(k_n, k_m) .
\end{eqnarray}
$C_3(n)$ represents the number of ways of summing the squares of three integers to equal $n$.

One obtains the energy levels and the composition of the energy eigenstates in finite volume by
solving the eigen-equation of the total isospin-zero Hamiltonian $\mathcal H^0=\mathcal
H^0_0+\mathcal H^0_{\rm int}$.  The results can be confronted with results from lattice QCD to
evaluate the merit of the model.

\section{Numerical results and discussion}\label{sec:NRD}

In this section, the cross sections of $K^-p$ and the eigen-energy spectrum for the $\Lambda(1405)$
are calculated in our two scenarios: one in which the $\Lambda(1405)$ is dynamically generated
purely from the $\pi \Sigma$, $\bar K N$, $\eta \Lambda$ and $K \Xi$ interactions, and one also
including a bare-baryon basis state to accommodate a three-quark configuration carrying the quantum
numbers of the $\Lambda(1405)$.  The poles for the $\Lambda(1405)$ resonance are determined, and
the associated structure is examined.

First, we obtain the couplings of the Hamiltonian field theory by fitting the cross sections of
$K^-p\to K^-p$, $K^-p\to\bar K^0n$, $K^-p\to \pi^-\Sigma^+$, $K^-p\to \pi^0\Sigma^0$, $K^-p\to
\pi^+\Sigma^-$, and $K^-p\to \pi^0 \Lambda$ at infinite volume.  With these couplings, the
eigen-energy levels can be calculated at finite volume. We compare them with the lattice QCD results and
discuss the structure of the $\Lambda(1405)$.

\subsection{Cross sections and poles} \label{subsec:ExpCSP}

Interactions in both the $I=0$ and $I=1$ channels contribute to the cross sections of
$K^-p$. Since the aim of this work is to study the $\Lambda(1405)$ at $I=0$, we include
$\pi\Sigma$, $\bar K N$, $\eta \Lambda$, and $K \Xi$ channels for $I=0$, while we only include
$\pi\Sigma$, $\bar K N$, and $\pi\Lambda$ channels for $I=1$.

\begin{figure}[t]
\begin{center}
\subfigure[\ $K^-p\to K^-p$]{\scalebox{0.24}{\includegraphics{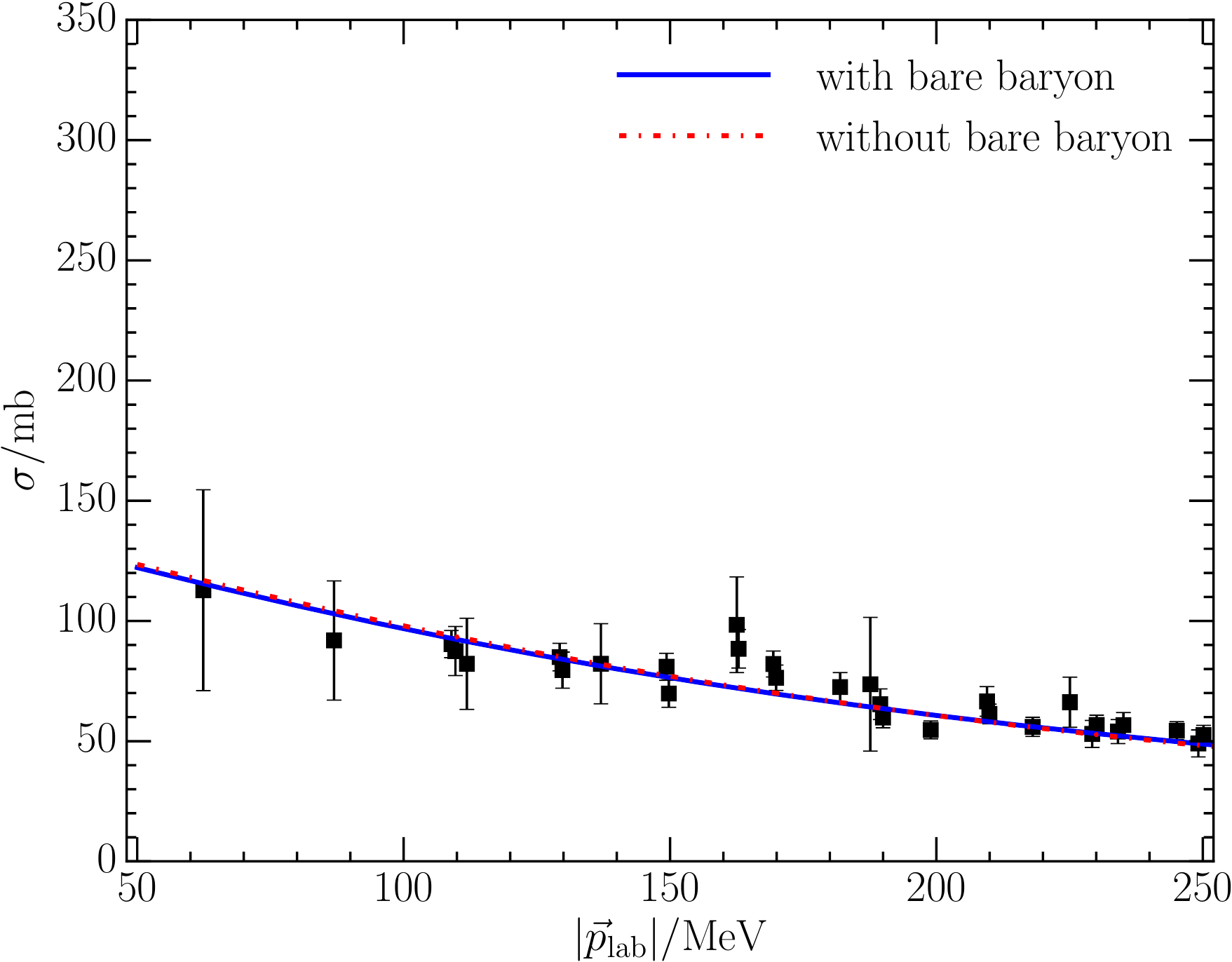}}}
\subfigure[\ $K^-p\to\bar K^0n$]{\scalebox{0.24}{\includegraphics{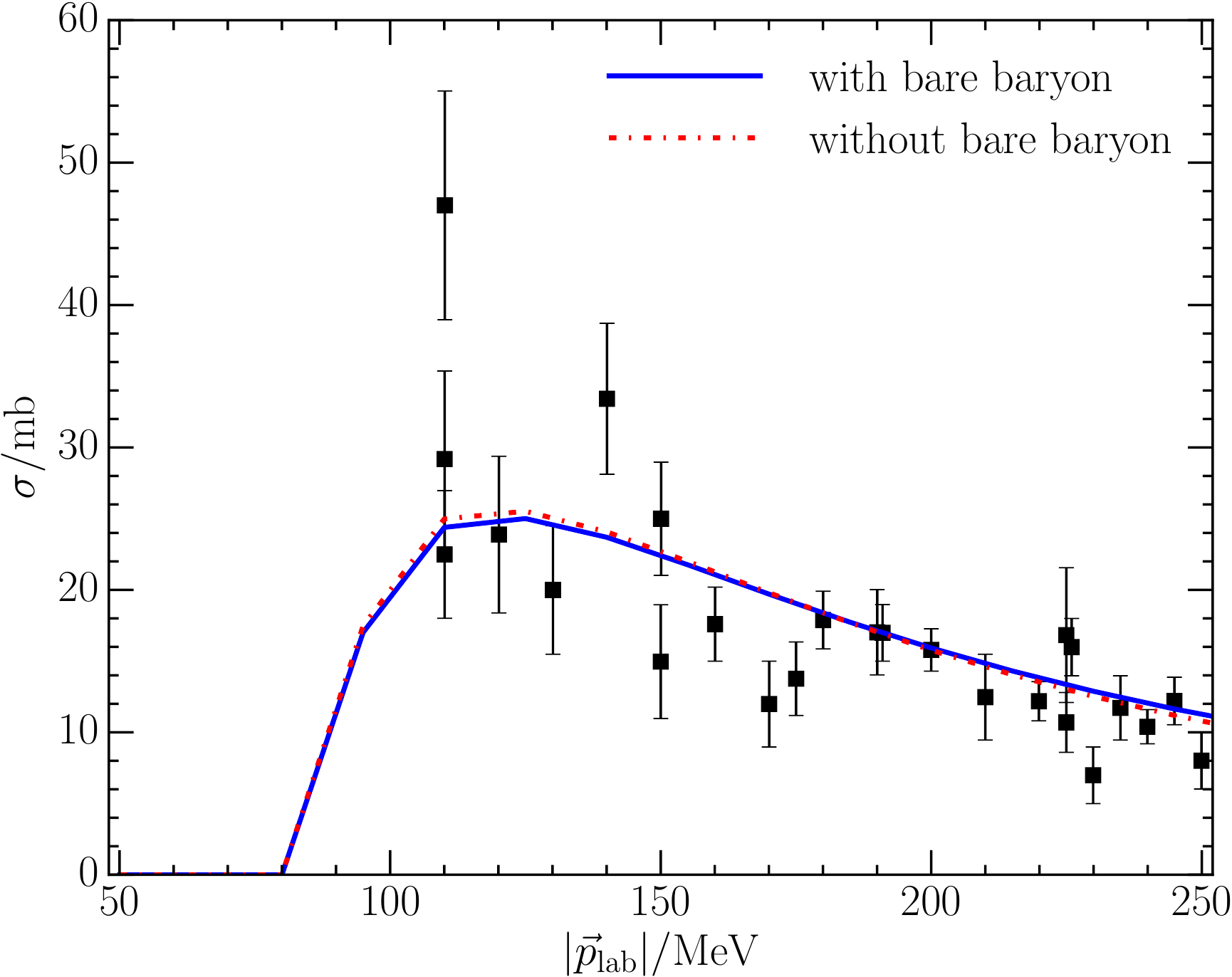}}}
\subfigure[\ $K^-p\to \pi^-\Sigma^+$]{\scalebox{0.24}{\includegraphics{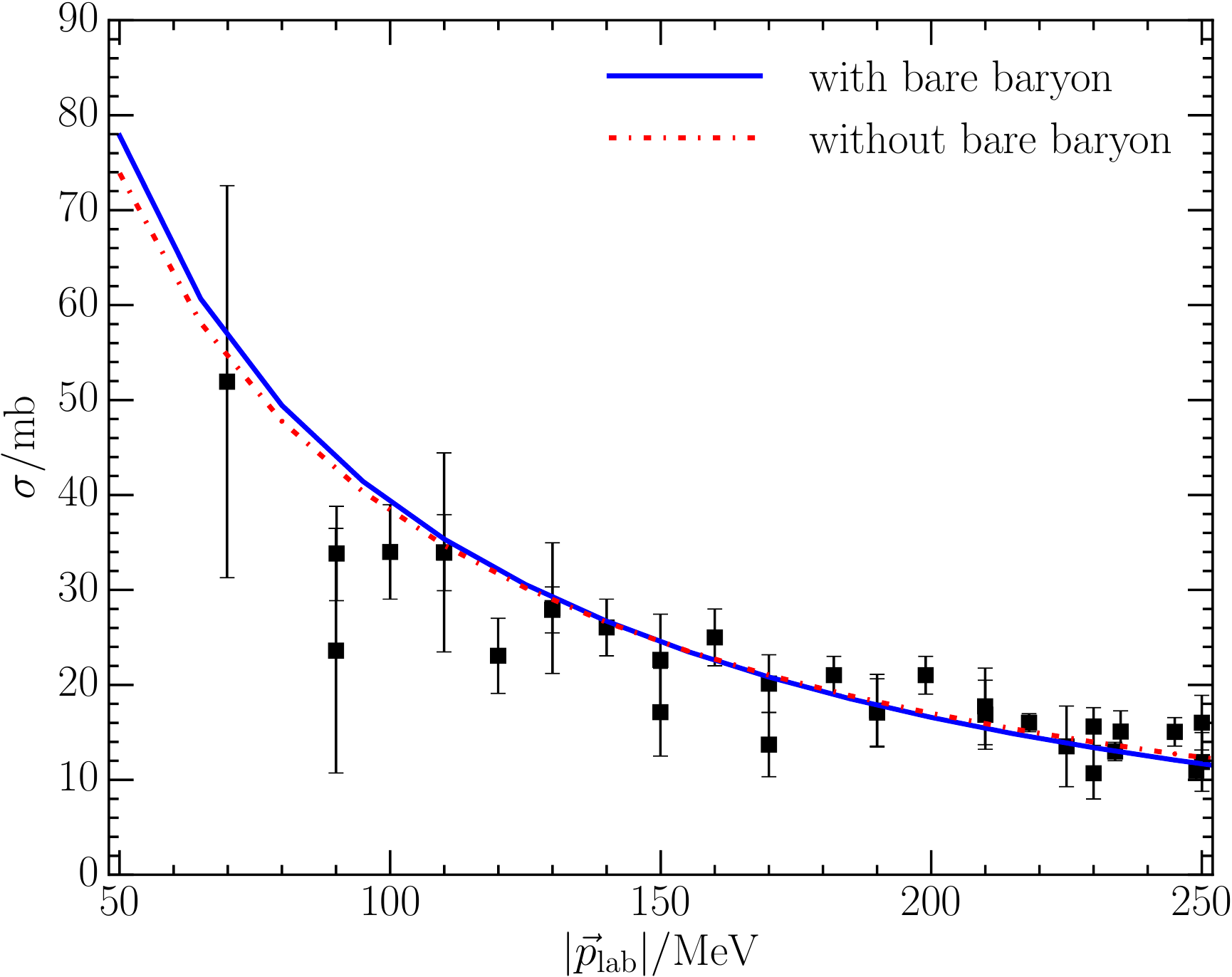}}}
\subfigure[\ $K^-p\to \pi^0\Sigma^0$]{\scalebox{0.24}{\includegraphics{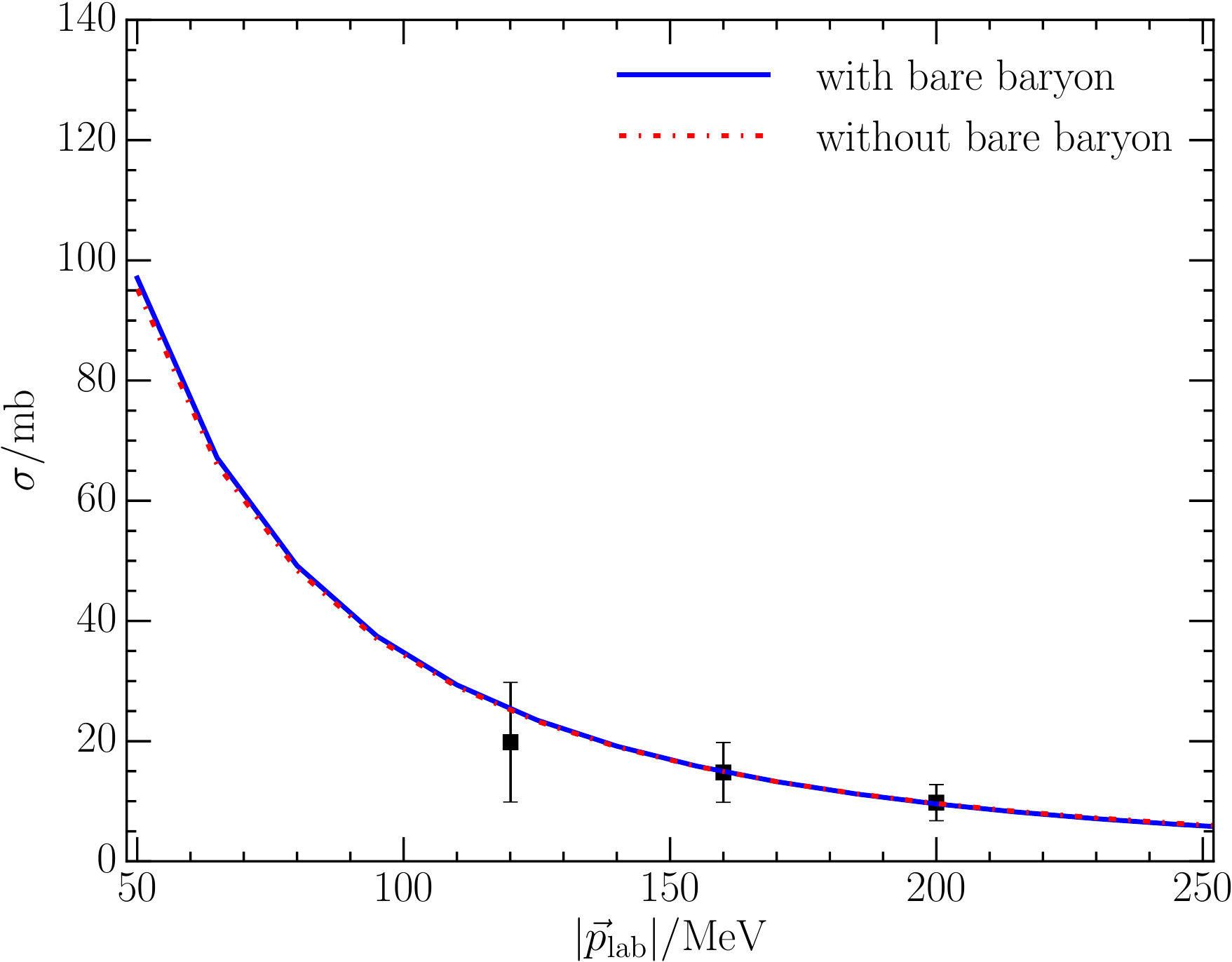}}}
\subfigure[\ $K^-p\to \pi^+\Sigma^-$]{\scalebox{0.24}{\includegraphics{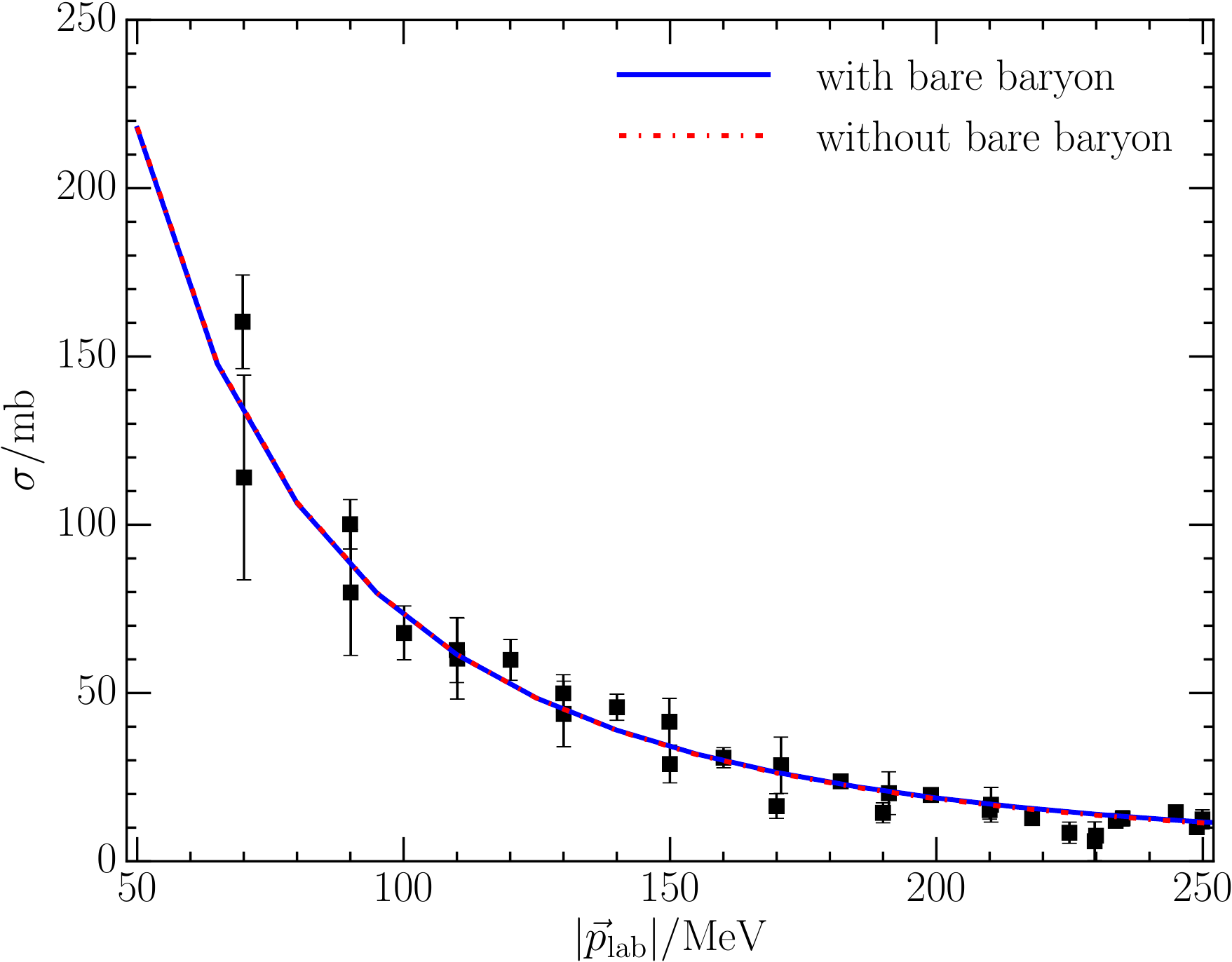}}}
\subfigure[\ $K^-p\to \pi^0 \Lambda$]{\scalebox{0.24}{\includegraphics{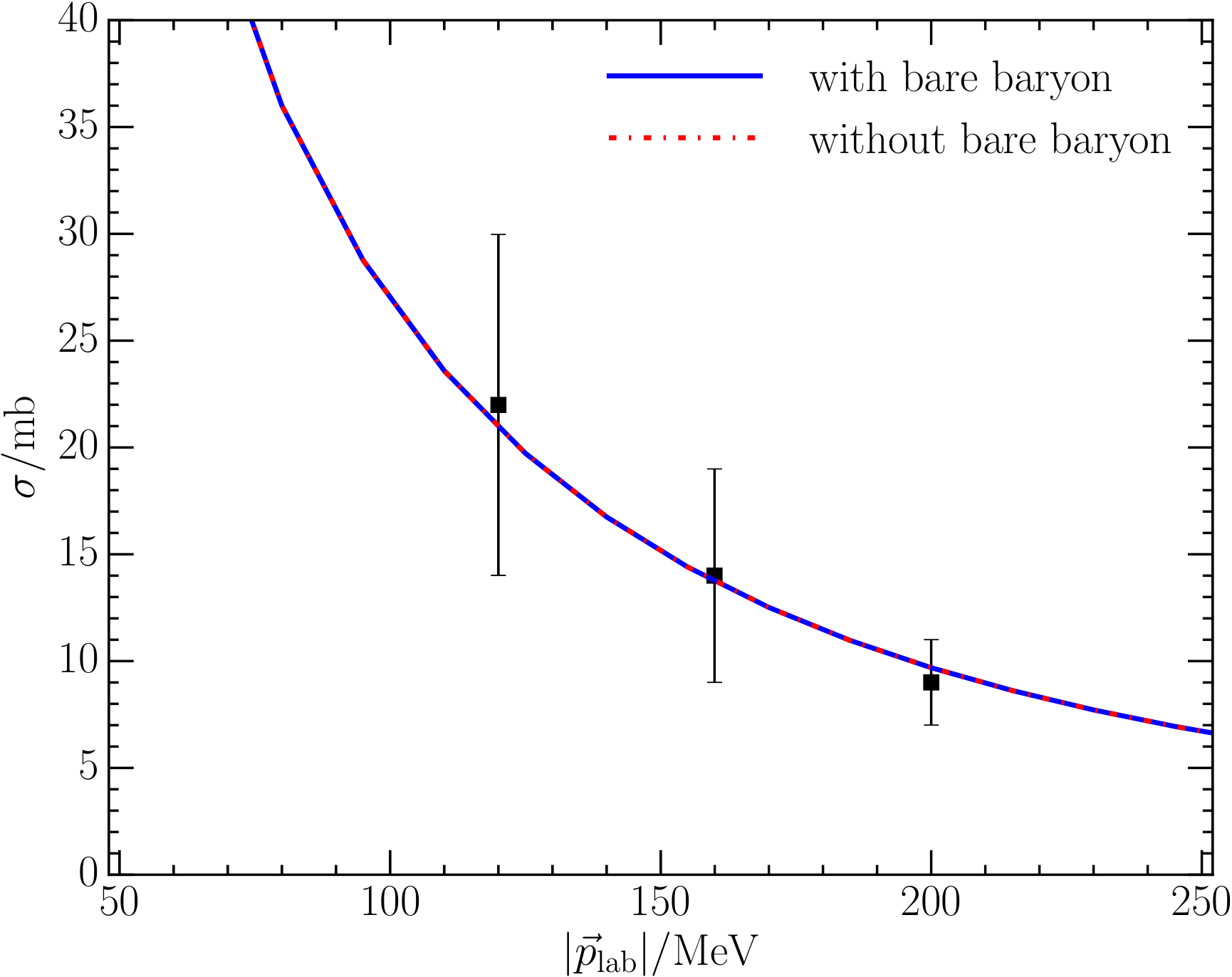}}}
\end{center} 
\caption{Experimental data and our fits to the cross sections of $K^- p$.  The solid lines are for
  our scenario with a bare-baryon component included in the $I=0$ channel, and the dashed lines
  represent the results without a bare-baryon component.  The experimental data are from
  Refs.~\cite{Abrams1965,Sakitt1965,Kim1965,Csejthey-Barth1965,Mast1976,Bangerter1981,Ciborowski1982,Evans1983}.}
\label{fig:CS}
\end{figure}

Since the threshold of $\eta \Lambda$ and $K \Xi$ are far away from the energy region of
experimental data, the cross sections are not very sensitive to their couplings. Therefore,
we set $ g_{\bar K N, \eta \Lambda}^0 $, $ g_{\pi\Sigma, K \Xi}^0 $, $ g_{\eta\Lambda, K \Xi}^0 $,
and $ g_{K \Xi, K \Xi}^0 $ at their ${\rm SU}_f(3)$-limit couplings but with one global adjustable
constant $g_0$,
\begin{eqnarray}
&&g_{\bar K N, \eta \Lambda}^0=-3/\sqrt{2} g_0, \quad
g_{\pi\Sigma, K \Xi}^0=-\sqrt{3/2} g_0,\nonumber\\
&&g_{\eta\Lambda, K \Xi}^0=3/\sqrt{2} g_0, \qquad
g_{K \Xi, K \Xi}^0=-3 g_0.
\end{eqnarray}
 
Comparing our two scenarios, the difference lies mainly in the $I=0$ channel where a bare
baryon can be included or omitted.  We first fit the cross sections in the scenario without a bare
baryon. After that, we leave the couplings in the channel $I=1$ fixed, and adjust those in the $I=0$
channel when incorporating a bare baryon contribution.

Just with limited experimental data for cross sections, we obtained a bare mass which can generate a pole close to that of $\Lambda(1670)$. The mass is far away from the fit energy region, and the properties of the bare state suffer from large uncertainties. In addition to the data for cross sections, we also fit the two masses from CSSM group at the largest two pion masses and make the pole in the infinite volume close to $(1670\pm 10)-(18\pm 7)i$ MeV at the same time in the second scenario.

The results of our fits to the cross sections with Eq. (\ref{eq:sigma}) are illustrated in
Fig.~\ref{fig:CS}.  The cross sections are described well, regardless of whether a bare baryon
contribution is introduced in the $I=0$ channel or not.  The fit parameters are provided in Table
\ref{tab:Para}.

Two poles are found for the $\Lambda(1405)$ in both scenarios.  The pole positions are consistent
with results from other groups, briefly reviewed in Table \ref{tab:Oth}. The real parts of the
poles are very close to the thresholds of $\bar K N$ and $\pi \Sigma$.

In our scenario without the bare state, we cannot find a pole for $\Lambda(1670)$. However, in the scenario with the bare baryon, we can find a pole corresponding to $\Lambda(1670)$ at $1660-30i$ MeV. Our result provides a possible candidate for $\Lambda(1670)$ which is mainly a bare state in our model.

The small differences in the $K^- p$ cross sections and the $\Lambda(1405)$ pole positions between
the two scenarios indicate that the $\Lambda(1405)$ contains little of the bare baryon component
at infinite volume.

\begin{table}[t]
\caption{Parameters constrained in our fits to cross sections of $K^- p$ and the pole positions
  obtained with these fit parameters in our two scenarios: one in which the $\Lambda(1405)$ is
  dynamically generated purely from the $\pi \Sigma$, $\bar K N$, $\eta \Lambda$ and $K \Xi$
  interactions (No $|B_0\rangle$), and one also including a bare-baryon basis state to accommodate
  a three-quark configuration carrying the quantum numbers of the $\Lambda(1405)$ (With
  $|B_0\rangle$).  The underlined entries indicate they are fixed in performing the fit.}
\label{tab:Para}
\begin{ruledtabular}
\begin{tabular}{lrr}
\noalign{\smallskip}
Coupling                                         & No $|B_0\rangle$  & With $|B_0\rangle$\\
\noalign{\smallskip}
\hline
\noalign{\smallskip}
$ g_{\pi\Sigma,\pi\Sigma}^0                 $    & $-1.77$ &$-1.59$                    \\
$ g_{\bar K N,\bar K N}^0                      $ & $-2.14$    & $-1.78$                    \\
$ g_{\bar K N,\pi\Sigma}^0          $            & 0.78      & 0.89                     \\
$ g_{\bar K N, \eta \Lambda}^0 $   &$-0.42$  &$-0.97$ \\
$ g_{\pi\Sigma, K \Xi}^0 $        &$-0.24$  &$-0.56$\\
$ g_{\eta\Lambda, K \Xi}^0 $   &0.42  &0.97\\
$ g_{K \Xi, K \Xi}^0 $    &$-0.60$  &$-1.37$\\
\noalign{\smallskip}
\hline
\noalign{\smallskip}
$ g_{\pi\Sigma, B_0}^0                 $ &- &0.13                    \\
$ g_{\bar K N, B_0}^0                      $ &- & 0.16                    \\
$g_{\eta\Lambda}^0$&-  &$-0.18$\\
 $g_{K \Xi}^0$ &-  &$-0.09$\\
$m_B^0/{\rm MeV}$ &-    &1740\\
\noalign{\smallskip}
\hline
\noalign{\smallskip}
$ g_{\pi\Sigma,\pi\Sigma}^1                 $  & $-0.14$     & \underline{$-0.14$}                  \\
$ g_{\bar K N,\bar K N}^1                      $  &$-0.06$    & \underline{$-0.06$}                   \\
$ g_{\bar K N,\pi\Sigma}^1           $          &1.36 &\underline{1.36}                     \\
$ g_{\bar K N,\pi\Lambda}^1           $    & 0.96 &\underline{0.96}\\
\noalign{\smallskip}
\hline
\noalign{\smallskip}
$\chi^2$ (120 data)    &166&177\\
\noalign{\smallskip}
\hline
\noalign{\smallskip}
pole 1  (MeV)                                    & $1428-23 \, i$  &$1429-22 \, i$               \\
pole 2 (MeV)                           &$1333-85 \, i$  &$1338-89 \, i$\\
\noalign{\smallskip}
\end{tabular}
\end{ruledtabular}
\end{table}

\begin{table}[tb]
\caption{Pole positions for the $\Lambda(1405)$ in various approaches.}\label{tab:Oth}
\begin{ruledtabular}
\begin{tabular}{lll}
Approach    & Pole 1 (MeV)  & Pole 2 (MeV)\\
\noalign{\smallskip}
\hline
\noalign{\smallskip}
Refs.~\cite{Valderrama2012,Ikeda2012} & $1424^{+7}_{-23}-i\, 26^{+3}_{-14}$ & $1381^{+18}_{-6}-i\,81^{+19}_{-8}$\\
Ref.~\cite{Guo2013} Fit I &$1417^{+4}_{-4}-i\, 24^{+7}_{-4}$ & $1436^{+14}_{-10}-i\,126^{+24}_{-28}$\\
Ref.~\cite{Guo2013} Fit II &$1421^{+3}_{-2}-i\, 19^{+8}_{-5}$ & $1388^{+9}_{-9}-i\,114^{+24}_{-25}$\\
Ref.~\cite{Mai2015} solution \#2 &$1434^{+2}_{-2}-i\, 10^{+2}_{-1}$ & $1330^{+4}_{-5}-i\,56^{+17}_{-11}$\\
Ref.~\cite{Mai2015} solution \#4 &$1429^{+8}_{-7}-i\, 12^{+2}_{-3}$ & $1325^{+15}_{-15}-i\,90^{+12}_{-18}$\\
This work &$1430-i\,22$ &$1338-i\,89$\\
\end{tabular}
\end{ruledtabular}
\end{table}

\begin{figure}[tb]
\begin{center}
\includegraphics[width=0.47\textwidth]{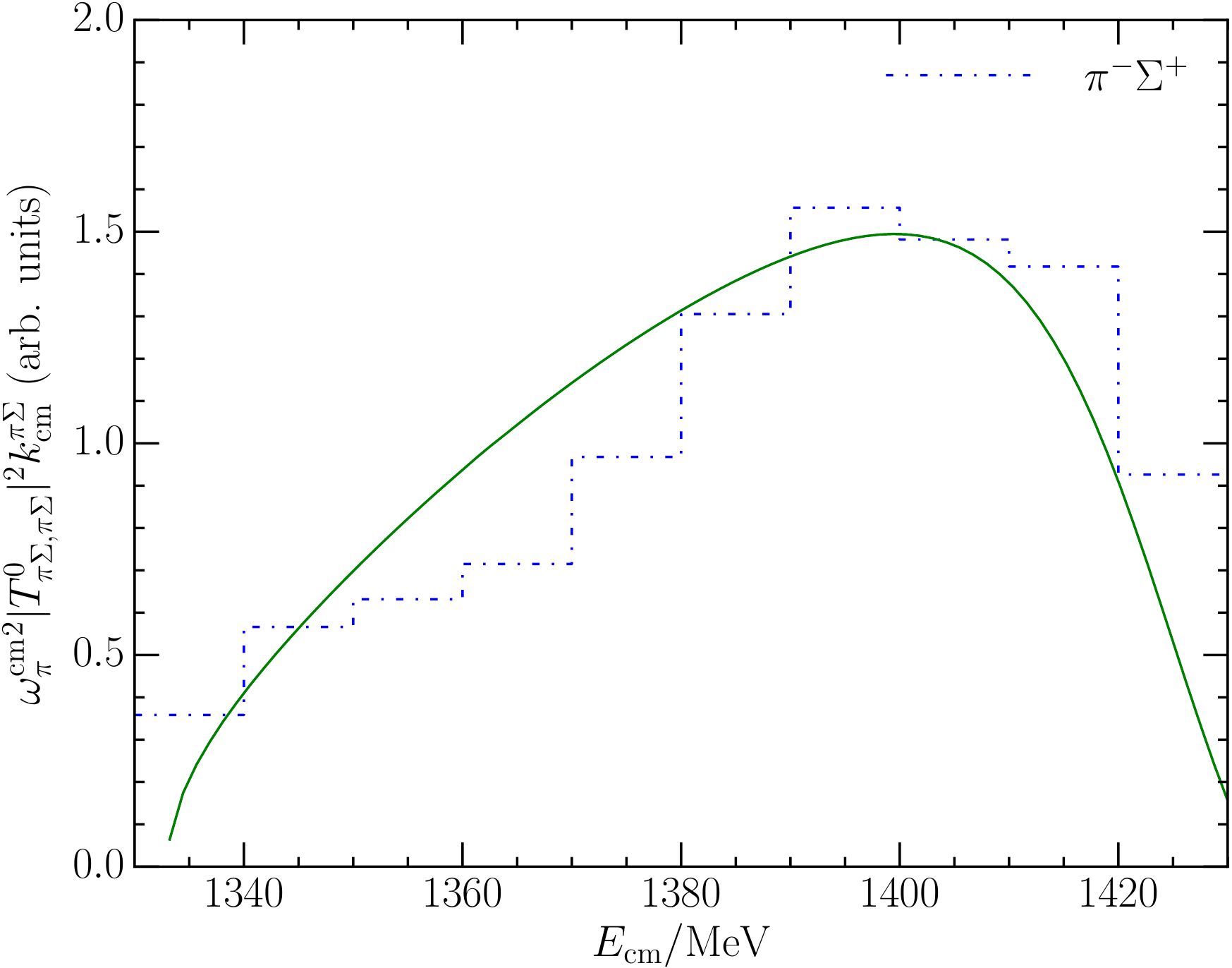}
\end{center}
\caption{{\bf Colour online:}
The $\pi\Sigma$ invariant mass distribution. 
The solid (green) curve is calculated from the scenario with the bare-baryon basis state and the
dashed (blue) histogram illustrates the experimental data from Ref.~\cite{Hemingway:1984pz}. 
}
\label{fig:massSpec}
\end{figure}

To explore the shape of the $\Lambda$(1405), we show the $\pi\Sigma$ invariant mass distribution in
Fig.~\ref{fig:massSpec}.  Here, the $y$-axis represents ${\omega_\pi^{\rm cm}}^2\, |T_{\pi \Sigma,
  \pi\Sigma}^0|^2 \, k_{\rm cm}^{\pi \Sigma}$
and the $x$-axis indicates the $\pi\Sigma$ centre-of-mass energy in units of MeV.  The leading
factor ${\omega_\pi^{\rm cm}}^2$ is due to the convention of $T$.  The solid (green) line is
calculated from our scenario with the bare baryon basis state.  Results for the scenario without a
bare baryon are very similar.  The dashed (blue) histogram illustrates the experimental data from
Ref.~\cite{Hemingway:1984pz}.  We note that the drop of the distribution from the peak is faster on
the right in our case and is consistent with the experimental data.

\subsection{Finite volume results for varying quark masses} \label{subsec:LattFV}

With the couplings determined and summarized in Table \ref{tab:Para}, we can proceed to determine
the finite-volume eigen-energy levels and associated components of the eigenstates by solving the
eigen-equation of the Hamiltonian $\mathcal H^0$ from Sec.~\ref{subsec:FVM}.  Of particular
interest is the impact of the bare-baryon basis state in the finite volume of the lattice over a
variety of pion masses.

To obtain results at larger pion masses, we need to know how the masses of the baryons and mesons
vary with the quark mass ($\propto m_\pi^2$).   For the bare mass, $m_B^0$, we use the linear assumption
\begin{equation}
m_B^0(m_{\pi}^2)=m_B^0|_{\rm phys.} +\alpha_B^0 \, (m_{\pi}^2-m_{\pi}^2|_{\rm phys.}) \, ,
\end{equation}
%
%
At larger quark masses, $\alpha_B^0$ should be approximately $\frac{2}{3} \alpha_{N(1535)}^0 = 0.51~{\rm GeV}^{-1}$
\cite{Liu2016a}.
For each of the masses $m_N(m_\pi^2)$, $m_\Sigma(m_\pi^2)$, $m_K^2(m_\pi^2)$ etc., we use a linear
interpolation between the corresponding lattice QCD results.

\subsection{Conventional Analysis}\label{sec:convAnalysis}

\begin{figure}[tb]
\begin{center}
\includegraphics[width=0.47\textwidth]{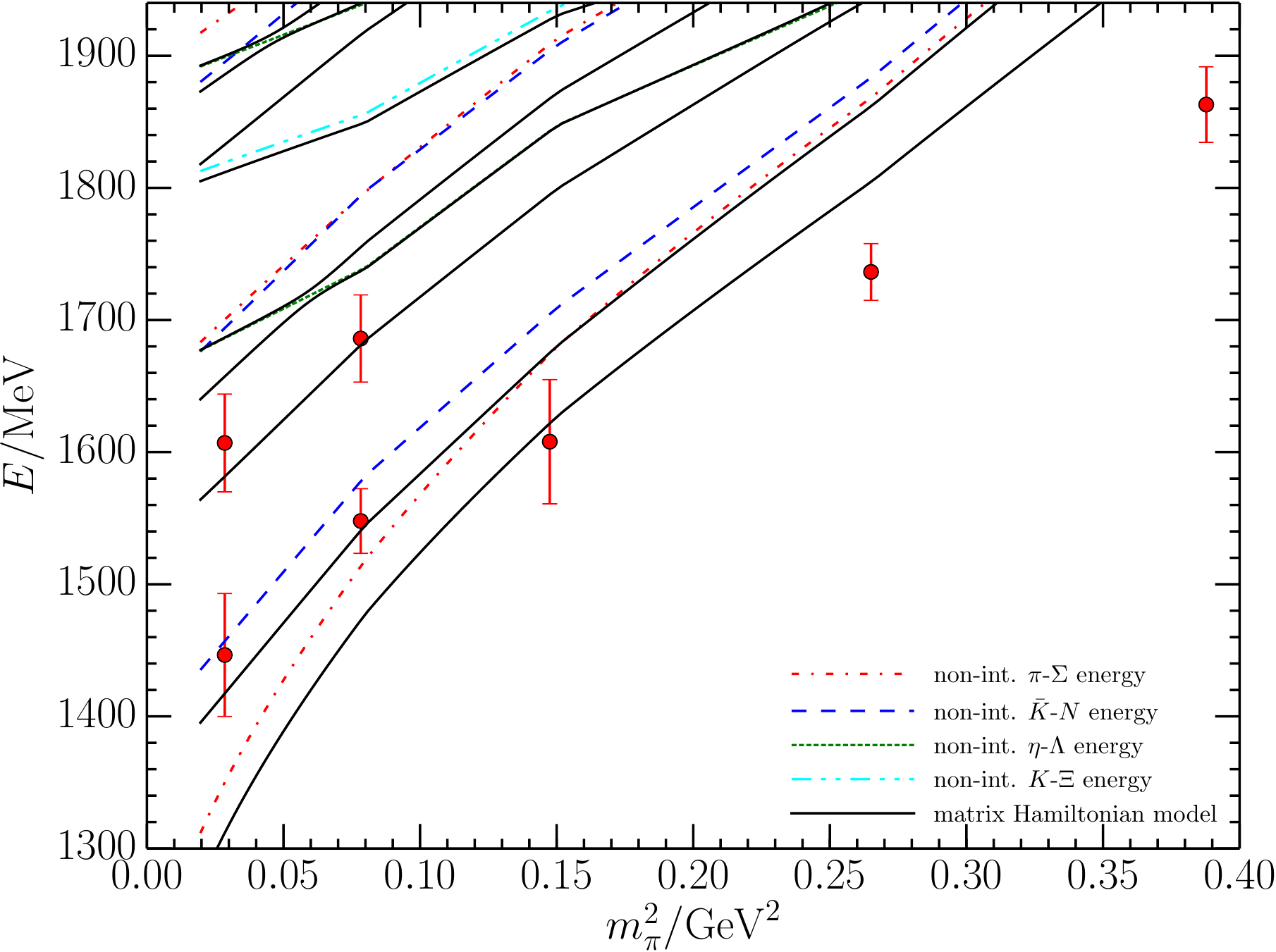}
\end{center}
\caption{{\bf Colour online:} The pion-mass dependence of the finite-volume energy eigenstates for
  the scenario without a bare-baryon basis state. The broken lines represent the non-interacting
  meson-baryon energies and the solid lines represent the spectrum derived from the matrix
  Hamiltonian model.  The lattice QCD results are from the CSSM \cite{Hall2015,Menadue2012}, as
  described in Table \ref{tab:CSSMlattice}.}
\label{fig:noBareSpec}
\end{figure}

The results of the model in the absence of a bare-baryon basis state are illustrated in
Fig.~\ref{fig:noBareSpec}.
Here we have used a linear interpolation between the CSSM results for the octet baryon masses.
We observe that while the model can fit the lattice results at low pion
masses, it fails at large pion masses. The results are very similar to those of
Ref.~\cite{Molina2015}, where the lattice results at large pion masses do not touch the curves
given by the model.
\begin{table}[tb]
\caption{The low-lying odd-parity $\Lambda$ masses provided by the CSSM group
  \cite{Hall2015,Menadue2012} with the strange-quark hopping parameter $\kappa_s=0.13665$ tuned to
  reproduce the physical Kaon mass \cite{Menadue2012}.  Values for $m_1$ are from
  eigenstate-projected correlators dominated by the flavour-singlet interpolator \cite{Hall2015}
  while values for $m_2$ are from projected correlators dominated by flavour-octet interpolators
  \cite{Menadue2012}.  Values, provided with reference to the pion mass, are in units of
  GeV.}\label{tab:CSSMlattice}
\begin{ruledtabular}
\begin{tabular}{cccccc}
$m_\pi$    &  0.6233(7)  & 0.5148(7)  &0.3890(10) &0.2834(6) &0.1742(26)\\
\noalign{\smallskip}
\hline
\noalign{\smallskip}
$m_1$      &  1.446(46) &1.548(24) &1.608(47) &1.736(21) &1.863(29)\\
$m_2$      &    - &               -            &    -            &1.686(33)& 1.607(37)\\
\end{tabular}
\end{ruledtabular}
\end{table}

\begin{figure*}[t]
\begin{center}
\subfigure[\ 1st eigenstate]{\scalebox{0.25}{\includegraphics{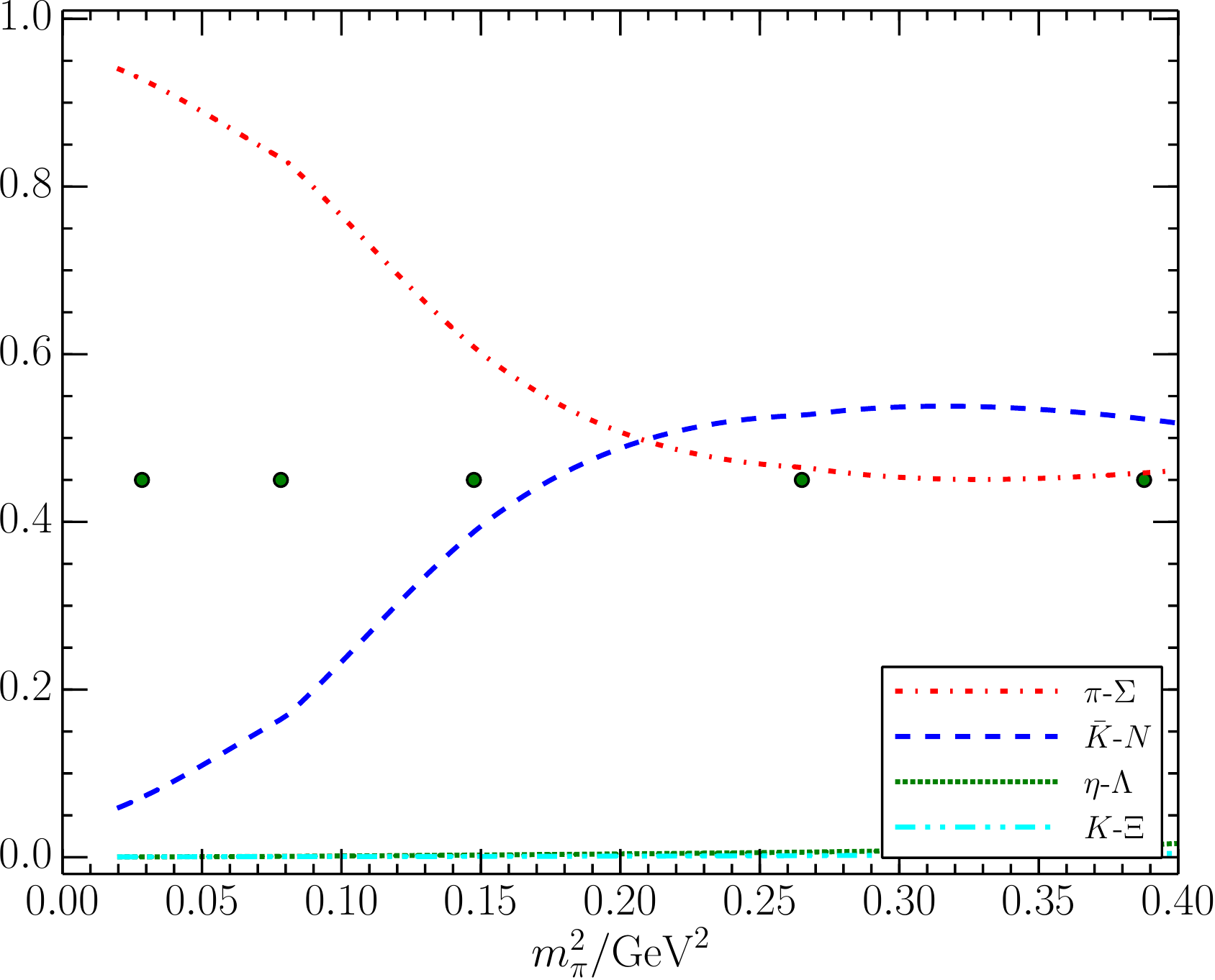}}}
\subfigure[\ 2nd eigenstate]{\scalebox{0.25}{\includegraphics{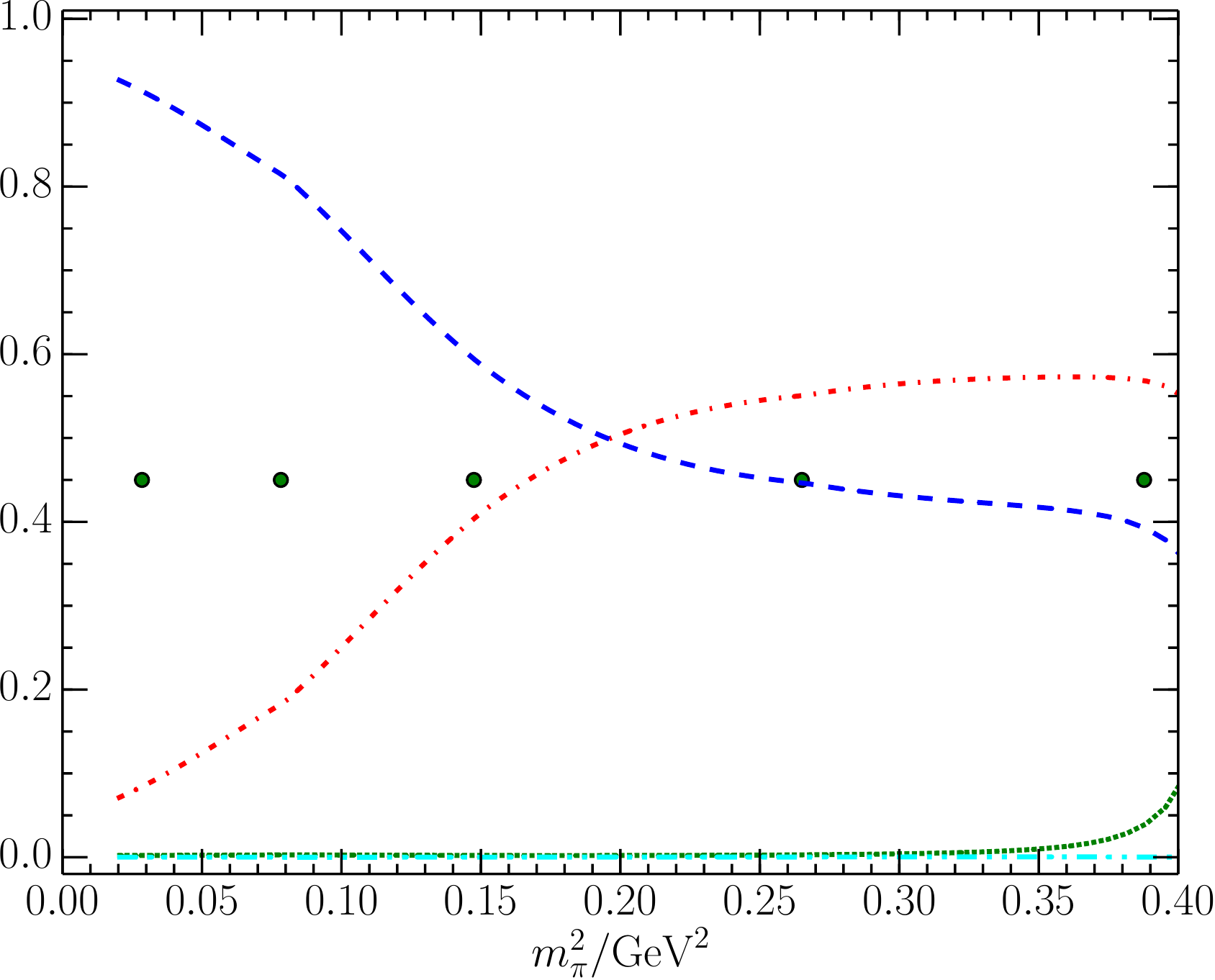}}}
\subfigure[\ 3rd eigenstate]{\scalebox{0.25}{\includegraphics{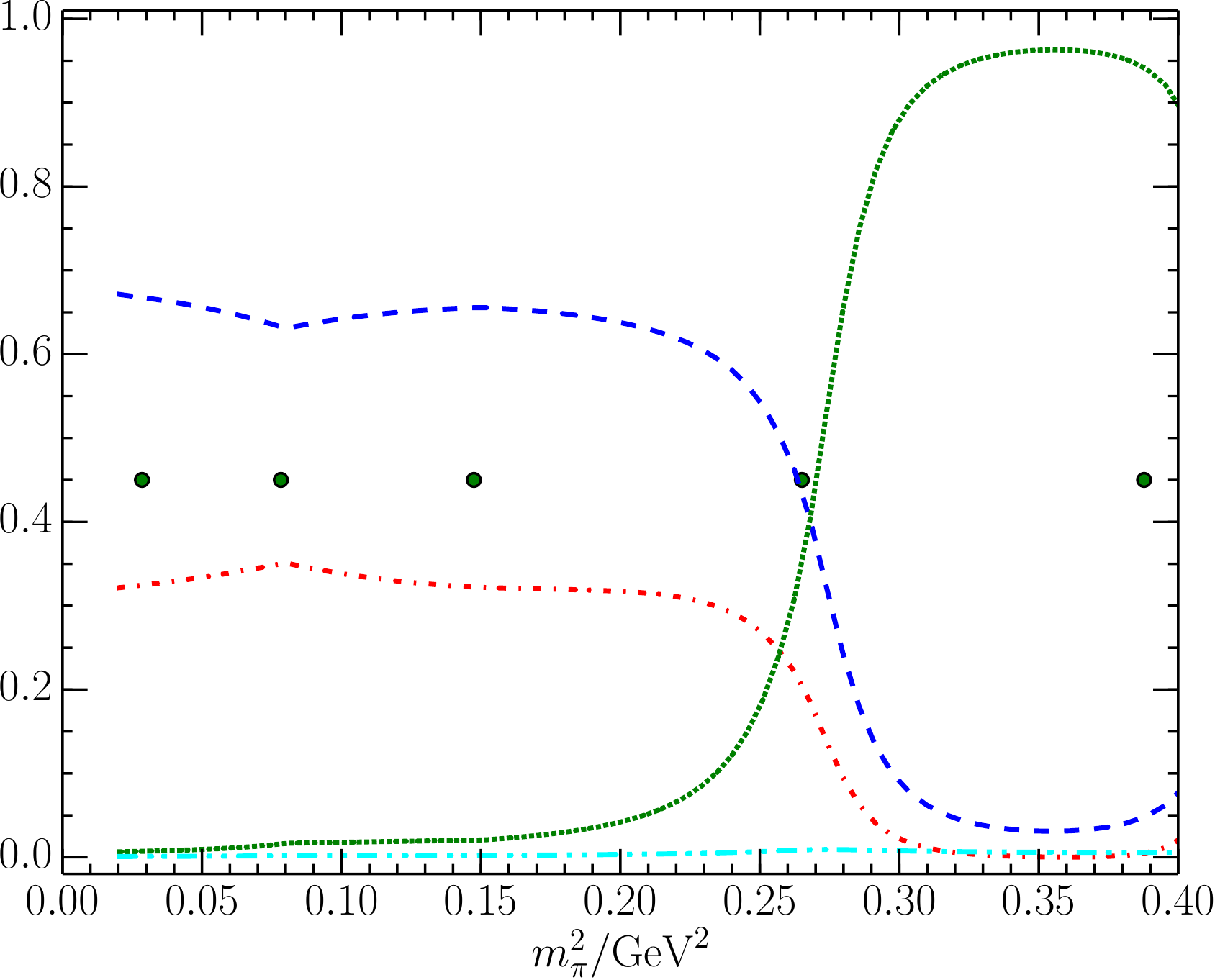}}}
\subfigure[\ 4th eigenstate]{\scalebox{0.25}{\includegraphics{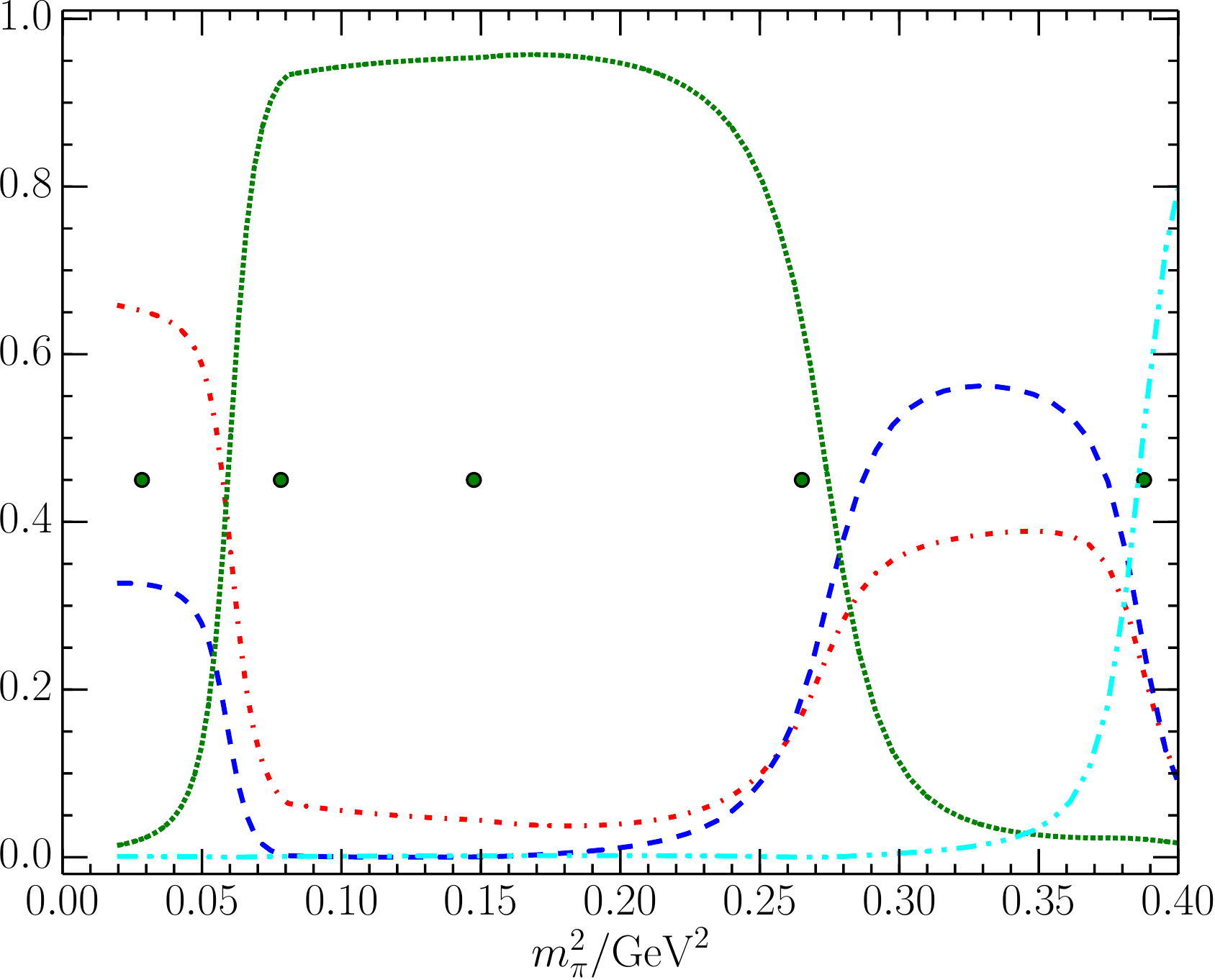}}}
\end{center} 
\caption{{\bf Colour online:} The pion-mass evolution of the Hamiltonian eigenvector components for
  the first four states observed in the scenario without a bare-baryon basis state.  Here all
  momenta for a particular meson-baryon channel have been summed to report the relative importance
  of the $\pi \Sigma$, $\bar K N$, $\eta \Lambda$, and $K \Xi$ channels. The (green) dots plotted
  horizontally at $y = 0.45$ indicate the positions of the five quark masses considered by the CSSM
  on a lattice volume with $L\simeq 2.90$ fm.}
\label{fig:NoBareComp}
\end{figure*}

The components of the eigenstates from the model without a bare-baryon basis state are presented in
Fig.~\ref{fig:NoBareComp}.  Panels~\ref{fig:NoBareComp}(a) and (b) reveal an avoided level crossing
in the low-lying $\pi \Sigma$ and $\bar K N$ dominated states.  At the lightest quark mass the
first eigenstate is composed mainly of $\pi \Sigma$ while the second eigenstate is dominated by the
$\bar K N$ component.  The third state is composed of a nontrivial mix of the $\pi \Sigma$ and
$\bar K N$ channels.

Only the second and third eigenstates are observed on the lattice.  
Consideration of the positions of the Hamiltonian model eigenstates relative to the dominant
non-interacting basis states can provide some insight into the reasons for this.  Both the $\pi
\Sigma$ and $\bar K N$ dominated eigenstates sit below the noninteracting energies in
Fig.~\ref{fig:noBareSpec} indicating significant attractive interactions.  The small Compton
wavelength of the kaon combined with attractive interactions with the nucleon could provide
significant clustering in the $\bar K N$ system.  Such clustering increases the probability of
finding the $\bar K$ next to the nucleon, thus increasing the overlap of the $\bar K N$-dominated
state with the local three-quark operators used to excite the state \cite{Menadue2012}.  Without
this strong attraction the overlap is volume, $V$, suppressed with the probability of finding the
meson next to the baryon $\propto 1/V$.  In the case of the $\pi \Sigma$-dominated state the large
Compton wavelength of the pion appears to reduce the level of clustering.  We will return to this
issue in the next Section including a bare-baryon basis state.

Beyond the third quark mass considered, the lattice QCD results depart from the eigenstates of the
Hamiltonian model.  At these pion masses, the $\Lambda(1405)$ has become a stable state lying lower
than the conventional $\pi \Sigma$ decay channel.  As for the nucleon, one expects a dominant role
for the simplest three-quark Fock-space component of the $\Lambda(1405)$ and the incorporation of a
bare-basis state in the Hamiltonian model will be essential to describing these results.  Drawing
on the results of Ref.~\cite{Liu2016} for the ground-state nucleon, one can anticipate that the
bare-baryon basis state will compose 80 to 90\% of the eigenvector components.  Therefore, we do
not trust the Hamiltonian model results of Fig.~\ref{fig:NoBareComp} at large quark masses.

\subsection{Inclusion of a bare-baryon basis state}\label{sec:BBanalysis}

\begin{figure}[t]
\begin{center}
\includegraphics[width=0.47\textwidth]{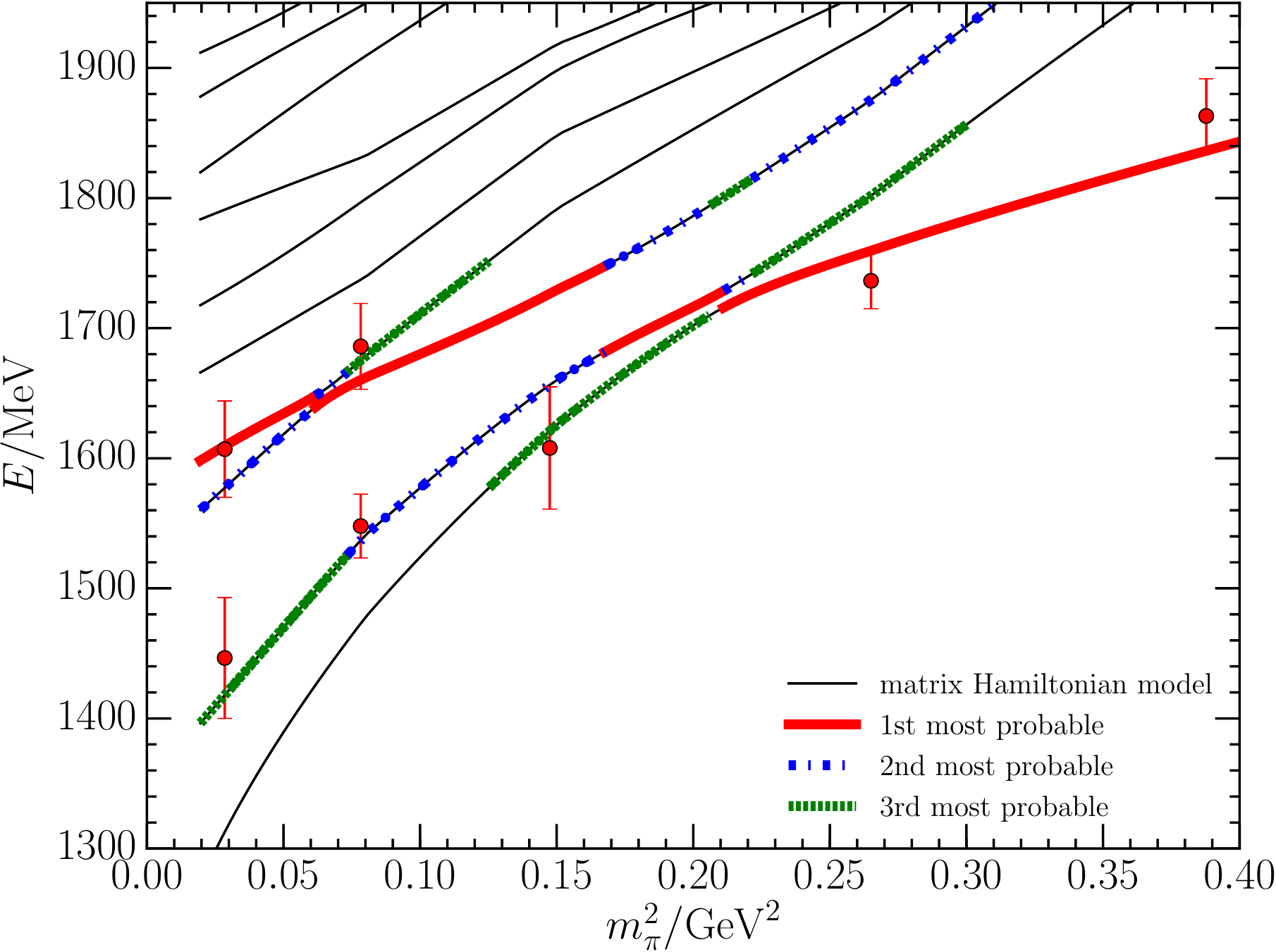}
\end{center}
\caption{{\bf Colour online:} The pion-mass dependence of the finite-volume energy eigenstates for
  the scenario including a bare-baryon basis state. The different line types and colours used in
  illustrating the energy levels indicate the strength of the bare basis state in the
  Hamiltonian-model eigenvector. The thick-solid (red), dashed (blue) and short-dashed (green)
  lines correspond to the first, second, and third strongest bare-state contributions, and
  therefore the most likely states to be observed with three-quark interpolating fields.}
\label{fig:BareSpec}
\end{figure}

\begin{figure}[t]
\begin{center}
\includegraphics[width=1.0\columnwidth]{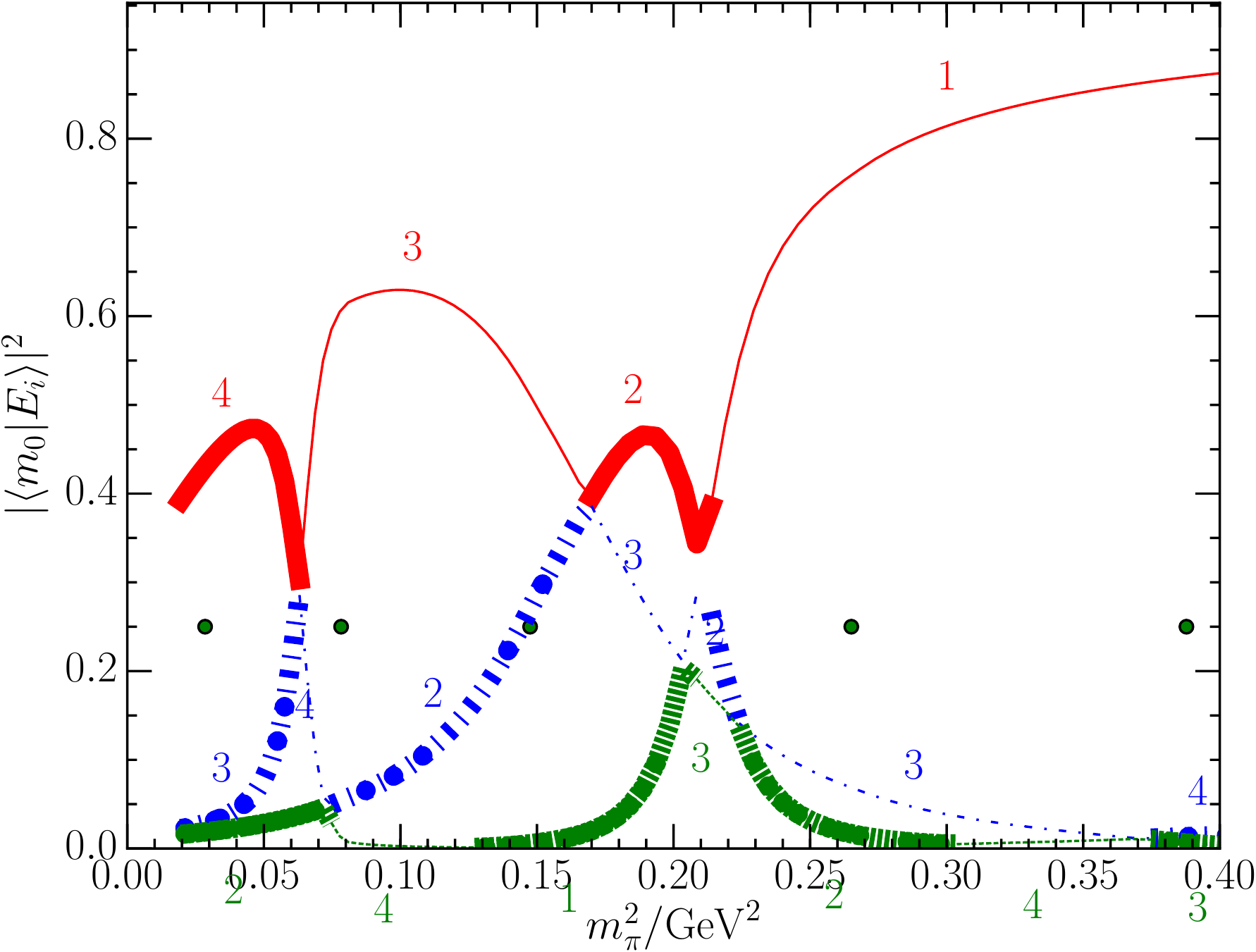}
\caption{{\bf Colour online:} The fraction of the bare-baryon basis state, $| m_0 \rangle$, in the
  Hamiltonian energy eigenstates $| E_i \rangle$ for the three low-lying states having the largest
  bare-state contribution.
  States are labeled by the energy-eigenstate integers $i$ indicated next
  to the curves.
  The dark-green dots plotted at $y = 0.25$ indicate the positions of the five quark masses
  considered in the CSSM results.  While the line type and colour scheme matches that of
  Fig.~\ref{fig:BareSpec}, the thick and thin lines alternate to indicate a change in the energy
  eigenstate.}
\label{fig:BareRFrac}
\end{center} 
\end{figure}

The inclusion of a bare-baryon basis state resolves the aforementioned discrepancies.  The pion
mass dependence of the odd-parity $\Lambda$ spectrum incorporating the bare-baryon basis state is
presented in Fig.~\ref{fig:BareSpec}.  Figure~\ref{fig:BareRFrac} indicates the states receiving
the largest contributions from the bare basis state.

Because local three-quark interpolating operators were used in exciting the states on the lattice,
one would expect that the states containing a significant bare state component are easier to
observe in lattice QCD.  As a result, we label the low-lying states containing the largest
bare-state components by superposing thick (coloured) lines on them in Fig.~\ref{fig:BareSpec}.  In
a successful description, the lattice results would correspond to these labeled states.

For example, the integers next to the solid red curve in Fig.~\ref{fig:BareRFrac} indicate
that the most probable state to be observed in lattice QCD simulations with local three-quark
operators is the fourth eigenstate at the lightest quark mass considered, becoming the 
third eigenstate for the second and third quark masses with $0.06 \le m_\pi^2 \le 0.16$ GeV${}^2$.   
As $m_\pi^2$ continues to increase, the most probable state falls to the second eigenstate briefly,
before settling on the lowest-lying state at the largest quark masses considered.

\begin{figure}[t]
\begin{center}
\includegraphics[width=0.47\textwidth]{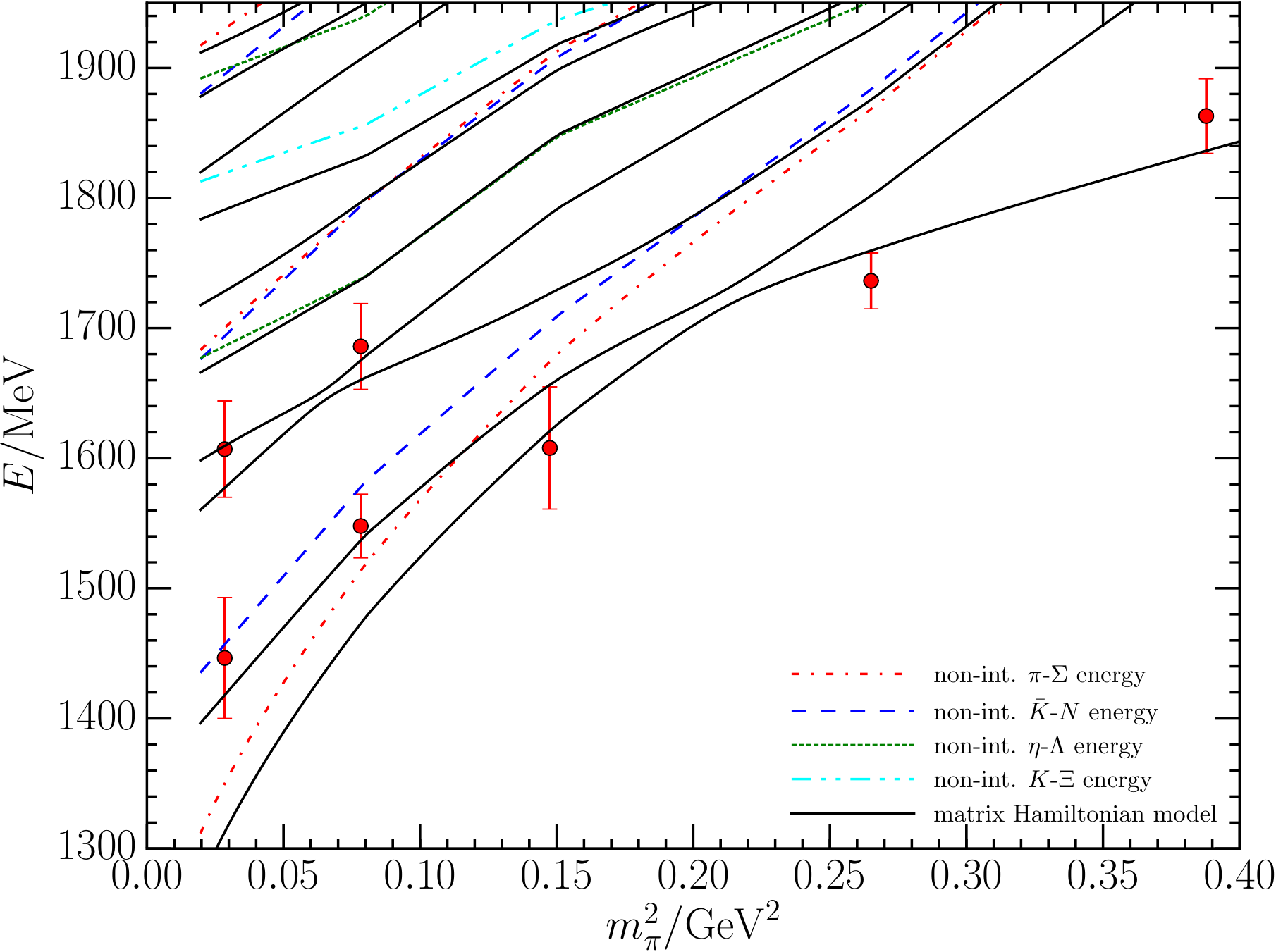}
\end{center}
\caption{{\bf Colour online:} The pion-mass dependence of the finite-volume energy eigenstates for
  the scenario including a bare-baryon basis state.  The broken lines represent the non-interacting
  meson-baryon energies and the solid lines represent the spectrum derived from the matrix
  Hamiltonian model.  
}
\label{fig:BareSpecDashed}
\end{figure}

Consideration of the three most probable states to be seen in the lattice QCD calculations is
sufficient to explain the states observed in the lattice QCD calculations.  At each mass, the
lowest-lying probable state(s) are observed.  The second excitations~\cite{Menadue2012}
observed on the lattice at the lightest two quark masses considered also agree with the energies of the
most probable states to be seen.

In contrast, the lowest-lying $\pi \Sigma$-dominated state at light quark masses has a negligible
bare-state component and therefore is not observed in the lattice QCD spectrum obtained with local
three-quark operators.  Instead, the lowest-lying lattice results correspond to the second
eigenstate which has both a bare state contribution and the benefit of clustering in the $\bar K N$
channel as discussed in Sec.~\ref{sec:convAnalysis}.  Figure~\ref{fig:BareSpecDashed} illustrates
the positions of the Hamiltonian model eigenstate energies relative to the non-interacting
meson-baryon basis-state energies, indicating attractive interactions.
Five-quark lattice operators with the momentum of both the $\pi$ and $\Sigma$ hadrons projected to
zero are expected to reveal the lowest-lying $\pi \Sigma$-dominated state predicted by the
Hamiltonian model.

\begin{figure*}[t]
\begin{center}
\subfigure[\ 1st eigenstate]{\scalebox{0.25}{\includegraphics{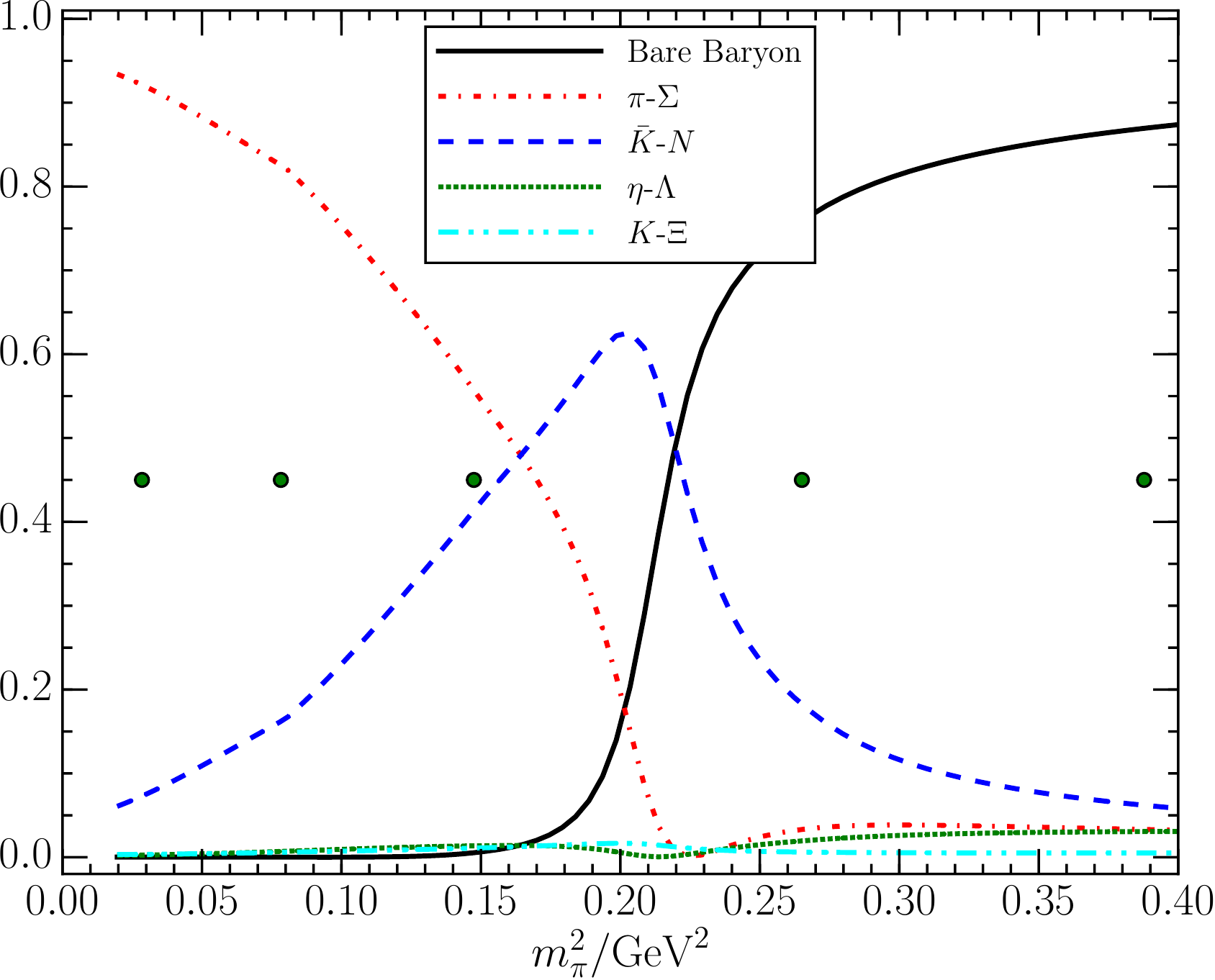}}}
\subfigure[\ 2nd eigenstate]{\scalebox{0.25}{\includegraphics{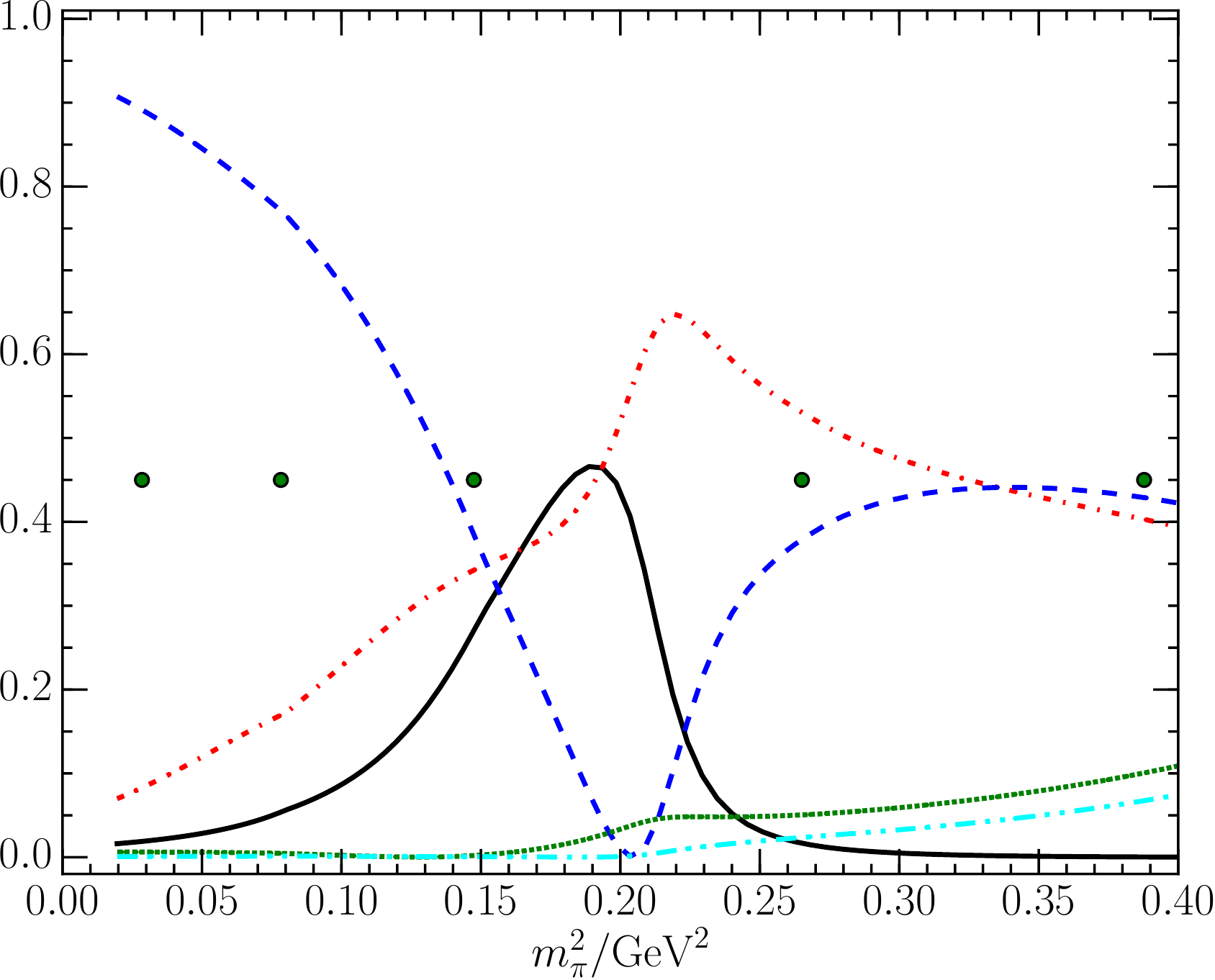}}}
\subfigure[\ 3rd eigenstate]{\scalebox{0.25}{\includegraphics{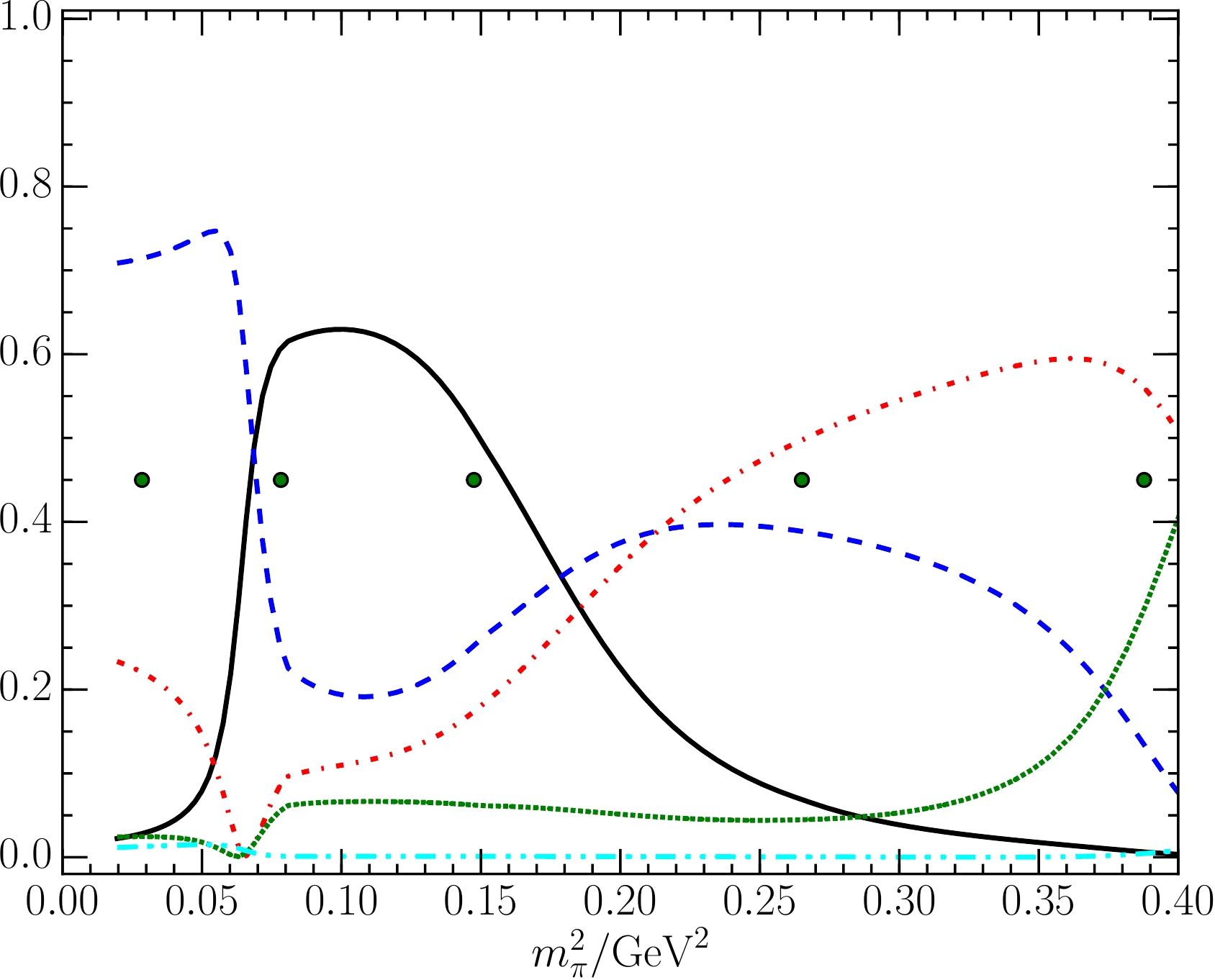}}}
\subfigure[\ 4th eigenstate]{\scalebox{0.25}{\includegraphics{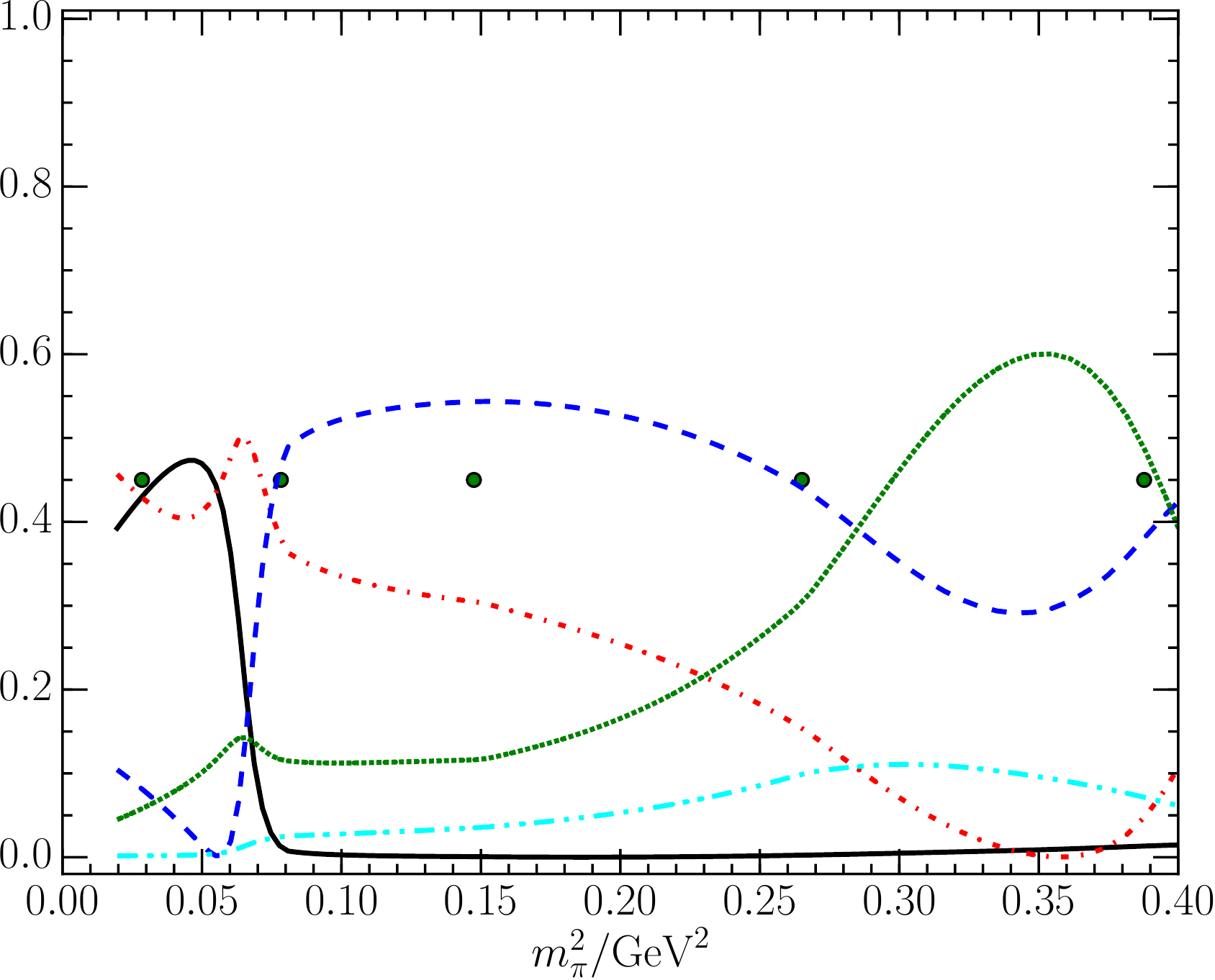}}}
\end{center} 
\caption{{\bf Colour online:} The pion-mass evolution of the Hamiltonian eigenvector components for
  the first four states observed in the scenario incorporating a bare-baryon basis state
  contribution.  Again, all momenta for a particular meson-baryon channel have been summed to
  report the relative importance of the meson-baryon channels. The (green) dots plotted
  horizontally at $y = 0.45$ indicate the positions of the five quark masses considered by the CSSM
  on a lattice volume with $L\simeq 2.90$ fm.}
\label{fig:BareComp}
\end{figure*}

The basis state components of the eigenstates for this scenario incorporating a bare-baryon basis
state are illustrated in Fig.~\ref{fig:BareComp}.  Considering Fig.~\ref{fig:BareComp}(a), one
observes that the first eigenstate is $\pi \Sigma$ dominated at small pion masses.  It transitions briefly
to a significant $\bar K N$ component and is eventually dominated by the bare-baryon basis state at
large pion masses.  
Comparing Figs.~\ref{fig:noBareSpec} and \ref{fig:BareSpec}, it is apparent that the bare
baryon is vital to describing the lattice QCD results for the $\Lambda(1405)$.

The uncertainty on the lattice QCD result at the middle quark mass considered ($m_\pi^2 = 0.15$
GeV${}^2$) is unusually large due to difficulty in identifying a plateau in the effective mass with
an acceptable $\chi^2_{\rm dof}$ at early Euclidean times.  Extensive Euclidean time evolution
isolated the lowest state in the spectrum at the expense of a larger uncertainty.  The origin of
the difficulty is now clear.  There are two nearby states in the spectrum at this quark mass, both
having significant overlap with the three-quark interpolating fields used.  Both the first and
second Hamiltonian eigenstates have large attractive $\bar K N$ components and both states have
nontrivial bare state components.  While the second state has a larger bare state component,
Euclidean time evolution will eventually favour the lower-lying state.  At moderate Euclidean
times, a superposition of states is encountered, accompanied by a large $\chi^2_{\rm dof}$ in the
single-state ansatz.  Further Euclidean time evolution favours the lower-lying state and the
single-state $\chi^2_{\rm dof}$ becomes acceptable.

At lighter quark masses, the first Hamiltonian eigenstate has a negligible bare-state component.
The second Hamiltonian eigenstate has both the attractive $\bar K N$ component and a nontrivial
bare-state component and is therefore seen on the lattice.  The fourth and third Hamiltonian model
eigenstates capture the largest bare-state components at the lightest and second-lightest quark
masses considered respectively and are associated with the lattice QCD eigenstates dominated by
SU(3)-flavour octet interpolating fields.

It is interesting to compare the spectra and structure observed herein at light quark masses with
the analysis of Ref.~\cite{Molina2015}.
Comparing Fig.~\ref{fig:BareSpecDashed} here with Fig.~1 of Ref.~\cite{Molina2015} for the
spectrum, both spectra commence with a bound state below the $\pi \Sigma$ threshold at small pion
masses. Consistently, the lowest lattice QCD results correspond to the second eigenstate for small
$m_\pi$.
The third eigenstate energies are both above 1500 MeV.  However, the third eigenstate energy
reported herein is 50 MeV larger at the physical pion mass.  At this energy, there is a desire to
consider experimental data at higher energies to better constrain the models and improve the
accuracy of the predictions.
In both works, four eigenstates are predicted below 1.6 MeV.

With regard to the composition of the states, we can compare Table III of Ref.~\cite{Molina2015}
with Fig.~\ref{fig:BareComp} in our work.  At the physical pion mass, both analyses indicate the
1st and 4th eigenstates are dominated by $\pi\Sigma$ basis states while the 2nd and 3rd eigenstates
are dominated by $\overline K N$ basis states.

Turning our attention to the quark-mass dependence of the spectrum, the isospin-zero bare-mass state
is associated with the lowest-lying state observed in lattice QCD calculations at large quark
masses.  However, as one moves away from the flavour-symmetric limit towards the light quark-mass
regime and flavour symmetry is broken, this bare mass becomes associated with the low-lying
flavour-octet dominated states.
As $m_\pi^2$ decreases, the first shift of the bare mass from state 1 to state 2 occurs at the
avoided level crossing of the $\pi \Sigma$ and $\Lambda(1405)$ at $m_\pi^2 = 0.21$ GeV${}^2$,
easily identified in Fig.~\ref{fig:BareSpec}.  Shortly thereafter, Hamiltonian-model eigenstate 1
becomes $\pi \Sigma$ dominated and the $\Lambda(1405)$ moves to the second eigenstate.
Moving to lighter quark masses, the bare mass shifts to the flavour-octet dominated states while
the flavour-singlet dominated $\Lambda(1405)$ evolves to become predominantly $\bar K N$, in accord
with the conclusions of Ref.~\cite{Hall2015}.  Its energy is around 1.446(46) GeV near the physical
pion mass.  From Fig.~\ref{fig:BareComp}(b), it is composed of about 90\% $\bar K N$, a few percent
$\pi \Sigma$, and a small amount of the bare-baryon basis state.

\section{Summary}\label{sec:sum}

We have studied the cross sections for $K^- p$ scattering at low energies using effective field theory.  We
considered two scenarios in constructing the basis states of our models: one in which the
$\Lambda(1405)$ is dynamically generated purely from the $\pi \Sigma$, $\bar K N$, $\eta \Lambda$
and $K \Xi$ interactions, and one also including a bare-baryon to accommodate a three-quark
configuration carrying the quantum numbers of the $\Lambda(1405)$.  Both scenarios produce
two-poles in the regime of the $\Lambda(1405)$ resonance, with values in accord with other
studies.  

With the parameters of the model constrained by the experimental data, Hamiltonian effective field
theory was used to calculate the finite-volume spectrum of states in our two scenarios and confront
lattice QCD data for the low-lying odd-parity $\Lambda$ spectrum in a finite volume with length
$L\sim 2.9$ fm.

At large quark masses, the bare state is vital to obtaining an accurate description of the
$\Lambda(1405)$.  Here the state is stable with a structure dominated by an 80 to 90\% bare state
component, similar to that for the ground state nucleon.  At smaller quark masses, the presence of
the bare-baryon basis state in the Hamiltonian model eigenvector explains which states are seen in
current lattice QCD calculations and which states are missed with local three-quark operators on
the lattice.

It is apparent that the nature of the $\Lambda(1405)$ changes dramatically as the light quark mass is varied. At large quark masses, the bare-baryon state is associated with the lowest-lying state observed in
lattice QCD calculations.  This state is excited by an interpolating field dominated by
SU(3)-flavour-singlet operators.  As one moves towards the light quark-mass regime, the bare
basis state becomes affiliated with the lattice QCD eigenstates excited by interpolating fields
dominated by flavour-octet operators.  After an avoided level crossing with the $\pi
\Sigma$-dominated state, the $\Lambda(1405)$ becomes the second state in the spectrum and evolves
to become a state dominated by $\bar K N$ components.  These results are consistent with the
earlier findings of Ref.~\cite{Hall2015} based on the strange quark contribution to the magnetic
form factor of the $\Lambda(1405)$.

Neither the cross sections for $K^- p$ scattering, nor the pole positions in the S-matrix are sensitive
to the bare-baryon basis state and this indicates that the physical resonance also has only a small
bare state component in its composition.  Together, these findings confirm that the $\Lambda(1405)$ is
predominantly a molecular $\bar K N$ bound state.

\begin{acknowledgments}
This research is supported by the Australian Research Council through the ARC Centre of Excellence
for Particle Physics at the Terascale (CE110001104), and through Grants No.\ LE160100051,
DP151103101 (A.W.T.), DP150103164, DP120104627 (D.B.L.). One of us (AWT) would also like to
acknowledge discussions with K. Tsushima during visits supported by CNPq, 313800/2014-6, and
400826/2014-3.
\end{acknowledgments}


\end{document}